\documentclass[11pt]{article}
\usepackage[utf8]{inputenc}

\usepackage{amsmath,amsfonts,graphicx}
\usepackage[page,header]{appendix}
\usepackage[square,numbers,sort&compress]{natbib}
\usepackage{fullpage}
\usepackage{lineno}
\usepackage{xcolor}
\usepackage{pdflscape}
\usepackage{titletoc}
\usepackage{hyperref}
\usepackage{multicol,multirow}
\usepackage{authblk}
\usepackage{pdfpages}

\newcommand{\hl}{}                    

\title{Multi-pathogen situational assessment and forecasting of respiratory disease in Aotearoa New Zealand}

\author[1,2]{Michael J. Plank}
\author[3,4]{Alys R. Young}
\author[3,5]{Katharine L. Senior}
\author[3,6]{Ruarai J. Tobin}
\author[7]{Mitchell O'Hara-Wild}
\author[8]{Fiona Callaghan}
\author[3,5]{Freya Shearer}
\author[3,6]{Oliver Eales}

\affil[1]{School of Mathematics and Statistics, University of Canterbury, Christchurch, New Zealand}

\affil[2]{Te P\=unaha Matatini, Auckland, New Zealand}

\affil[3]{Infectious Disease Dynamics Unit, Centre for Epidemiology and Biostatistics, Melbourne School of Population and Global Health, University of Melbourne, Melbourne, Victoria, Australia}

\affil[4]{College of Science and Engineering, James Cook University, Queensland, Australia}

\affil[5]{Infectious Disease Ecology and Modelling Team, The Kids Research Institute Australia, Perth, Western Australia, Australia}

\affil[6]{School of Mathematics and Statistics, University of Melbourne, Melbourne, Victoria, Australia}

\affil[7]{Monash University, Melbourne, Victoria, Australia}

\affil[8]{Public Health Agency, Manat\=u Hauora Ministry of Health, Wellington, New Zealand}

\date{}

\begin{document}

\newpage

\maketitle

\begin{abstract}
    Real-time analysis of epidemic trends and forecasts can help support public health planning and the response to seasonal respiratory disease. Here, we present two models that were used in a 2025 New Zealand winter situational assessment programme for three respiratory pathogens: SARS-CoV-2, influenza and respiratory syncytial virus (RSV). {\hl Data on SARS-CoV-2 were obtained from the national Covid-19 surveillance system; data on influenza and RSV were limited to a sentinel hospital surveillance programme.} Models were run weekly from May to October 2025 on these real-time disease surveillance data and provided a quantitative representation of the current epidemic trend, along with estimates of the epidemic growth rate and 28-day ahead forecasts of case incidence. Model results and interpretation were provided in weekly reports to public health partners as part of a trans-Tasman winter programme run by the Australia--Aotearoa Consortium for Epidemic Forecasting and Analytics (ACEFA). We compare in-season results that were included in these reports to a retrospective analysis of the complete data for the season. We conclude that real-time analyses performed reasonably well, and identify some areas for improvement in future winter situational assessment programmes.
\end{abstract}

\newpage 

\section{Introduction}

Analysis of current epidemic trends and near-term forecasts can contribute to situational awareness and support healthcare capacity planning \cite{moss2018epidemic,lutz2019applying}. This is particularly relevant during the winter respiratory illness season, when healthcare capacity is often stretched \cite{gozzi2025performance} due to overlapping respiratory pathogen epidemics. Real-time analysis and forecasting models can help anticipate and plan for increases in hospitalisations or patient load. 

Aggregate metrics, such as severe acute respiratory illness, or syndromic surveillance indicators such as influenza-like illness (ILI), are routinely used to monitor the progression of the respiratory illness season. However, these aggregate metrics can mask trends in the composite pathogens that contribute to the aggregate metric \cite{Eales2024-ro}. For example, a decreasing trend in an aggregate metric could mask a growing epidemic of a single component pathogen, and therefore lead to underestimation of future healthcare demand. Quantifying pathogen-specific trends (or strain-specific trends where relevant) provides richer and more valuable information than assessing trends in aggregate metrics alone \cite{Eales2025-cg}.     

Real-time trend analysis and forecasts have been applied to a variety of pathogens including influenza \cite{yang2015forecasting,zarebski2017model}, dengue \cite{reich2016challenges}, Zika \cite{chowell2016using}, Ebola \cite{funk2019assessing}, diphtheria \cite{finger2019real}. They can also play an important role in pandemic response by adding an early warning dimension to pandemic respiratory surveillance, and were used extensively during the COVID-19 pandemic \cite{cramer2022evaluation,moss2023forecasting,sherratt2023predictive,plank2024near}. 

In this paper, we describe a winter situational assessment programme for respiratory diseases, which was implemented for the first time in New Zealand in 2025. This programme was run by the Australia--Aotearoa Consortium for Epidemic Forecasting and Analytics (ACEFA) that oversees a trans-Tasman situational assessment programme, including the Australia--Aotearoa Forecasting Hub\cite{acefa_website}. This hub reflects the structure and activities of other infectious disease forecasting hubs around the world\cite{reich2016challenges}.

Our analysis focuses on two important components of situational assessment: real-time epidemic trend analysis and short-term forecasting. We present two separate models, developed by two independent teams, one for trend analysis and one for forecasting. Both of these models were applied to the same surveillance data streams representing SARS-CoV-2, influenza and respiratory syncytial virus (RSV) activity in New Zealand.
Our trend analysis model uses Bayesian P-splines \cite{wood2017psplines} to model a case or hospitalisation incidence time series as a smooth function of time. This allows us to estimate quantities such as the epidemic growth rate and the effective reproduction number. 

Our forecasting model is a latent state model with a renewal process for daily infection incidence \cite{fraser2007estimating} and a time-varying reproduction number \cite{abbott2020estimating}. The observed variables are daily case notifications and, where available, new daily hospital admissions (each of which is some fraction of infections, observed after a distributed time delay). The model is fitted to data using a particle filter (i.e., sequential Monte Carlo) approach \cite{steyn2025primer}.

For both models, we compare in-season results that were used in real-time reporting to public health partners against retrospective analysis of complete data for the 2025 season.

\section{Methods}

\subsection{Data}

We obtained two de-identified unit record datasets from Te Whatu Ora Health New Zealand. The first dataset contained all PCR-confirmed cases of SARS-CoV-2 in New Zealand since 26 February 2020, including admission date for patients admitted to hospital for treatment for COVID-19. {\hl This covers the national population of approximately 5.3 million people \cite{statsnz_pop_size}. During the study period, most PCR testing for SARS-CoV-2 was carried out in hospital settings, and all PCR results and related hospital admissions were reported to the national Public Health Agency and stored in a central data repository. The Public Health Agency categorises hospital admissions as SARS-CoV-2-related or incidental using two different sources of data. The first data source (the ``inpatient database'') is available in near-real-time but includes probable SARS-CoV-2 hospitalisations. The second data source (the ``national minimum dataset'') is based on the final clinical coding, which is more reliable but is delayed as hospital events are only required to be loaded into the dataset within 21 days after the month of discharge \cite{nmds}. We did not have access to the raw clinical datasets, only the categorisation updated in real-time by the Public Health Agency. This meant that revisions occurred to real-time counts due to hospital admissions being retrospectively added to the data, and due to admissions being later re-categorised as not SARS-CoV-2-related. However, the frequency and magnitude and these revisions were not known in advance. }

The second dataset contained all patients admitted overnight to {\hl one of the four hospitals in the Auckland and Counties Manukau Districts (Kidz First Children's Hospital, Starship Children's Hospital, Middlemore Hospital and Auckland City Hospital)} with severe acute respiratory infection (SARI) and who tested positive for either influenza or RSV since 2 January 2025. These data were collected by Public Health and Forensic Science as part of New Zealand's sentinel hospital-based SARI surveillance system \cite{huang2014implementing,phf2025acute}. 

{\hl It is important to note that the SARS-CoV-2 data are national, while the influenza and RSV data are from a regional sentinel surveillance programme. This covers the largest health region in New Zealand with a population of around 1.2 million people \cite{statsnz_pop_size} (i.e. around 23\% of the national total) with similar socioeconomic and demographic characteristics to the New Zealand population \cite{huang2014implementing}. The different population catchments preclude comparisons of the absolute number of cases between SARS-CoV-2 and either influenza or RSV, although comparisons in epidemic dynamics, such as the timing of the winter peak, are still possible with appropriate caveats. As with many clinical-based respiratory disease surveillance programs, the data relate to cases presenting to the healthcare system with acute respiratory symptoms. Additionally, since the surveillance is primarily hospital-based, cases are further skewed towards the severe end of the clinical spectrum and are not necessarily representative of acute respiratory infections in the wider community. }

Weekly updates to each dataset were provided throughout the 2025 winter season (15 rounds). For any given dataset, we refer to the latest date on which data were available as the origin date. The final update was received on 23 October 2025 and contained data up to 10 October 2025.
From each update of the unit record data, we calculated the daily total recorded number of laboratory-confirmed: SARS-CoV-2 case notifications, new SARS-CoV-2 hospital admissions, new influenza hospital admissions, and new RSV hospital admissions.

\subsection{Model for trend analysis}
We used a Bayesian P-spline model to quantify the past and current trends in two time series: i) SARS-CoV-2 case incidence and ii) SARS-CoV-2, influenza, and RSV hospitalisation incidence. The model has been described in full elsewhere \cite{Eales2025-cg}. In brief, we model the logarithm of the expected value of a time series (i.e., cases and hospitalisations) as a smooth function of time, $s(t)$, using a penalised-spline. A penalised-spline is a linear combination of $N$ basis splines of degree $n$, $B_{i,n}(t)$:
\begin{equation}
    s(t) = \sum_i^N b_iB_{i,n}(t).
\end{equation}
where $b_i$ are the basis spline coefficients. For any time series we define a system of third-order basis splines over regularly spaced knots. We set adjacent knots five days apart --- this has previously been shown to be a fine enough temporal resolution to capture dynamical effects of SARS-CoV-2 infection prevalence time series data \cite{Eales2022-yr}. The system of basis splines extends three knots either side of the time series to prevent edge effects. We assume a second-order random-walk prior on the basis spline coefficients:
\begin{equation}
    b_i = 2b_{i-1} - b_{i-2}+u_i,
\end{equation}
where
\begin{equation}
    u_i\sim N(0, \tau^2).
\end{equation}
This prior penalises changes in the first derivative of the smoothed function (i.e., changes to the growth rate), with the degree of penalisation set by the additional parameter $\tau$. We assume that time series data, $C(t)$, is negative binomially distributed:
\begin{equation}
    C(t) \sim \mathrm{NegBin}(e^{s(t)}, k), 
\end{equation}
with an additional parameter $k$ setting the over-dispersion of the data. 

We fit the model using a No-U-Turns Sampler (NUTS) \cite{Hoffman2014-mp} implemented in RStan \cite{Stan_Development_Team2020-sz}. {\hl For all model fits we confirmed that all parameters had an R-hat less than 1.02, and an effective sample size greater than 400.} When fitting the model to the SARS-CoV-2 case and hospitalisation time series, we assume uninformative constant prior distributions for the parameters $\tau$ and $k$. When fitting the model to influenza and RSV hospitalisation time series, we assume an informative normally distributed prior distribution for both $\tau$ and $k$ with mean and standard deviation taken as the mean and standard deviation of the posterior distributions of the model fit to SARS-CoV-2 hospitalisations (for the same week of reporting). This was done to allow the model to converge, when fitting to the shorter influenza and RSV time series. {\hl We also performed a sensitivity analysis by fitting the model (to the final dataset) assuming the mean of the informative prior distribution for $\tau$ was 0.5 or 2 times the posterior distribution of the model fit to SARS-CoV-2 hospitalisations.} 

{\hl For weekly reporting, the Bayesian P-spline models were fit to each dataset as received in real-time over the course of the season.} The Bayesian P-spline model was fit to the previous three years of data (up to the last day of data {\hl available in real-time}) for SARS-CoV-2 cases and hospitalisations. The model was fit to the entire continuous time series for influenza and RSV hospitalisations (first day of available continuous data was 2 January 2025{\hl, until last day of data available in real-time}). When fitting the model to SARS-CoV-2 case time series we additionally account for day-of-the-week effects in the time series (see \cite{Eales2025-cg}). {\hl As a sensitivity analysis, we also fit the Bayesian P-spline model to the final dataset for SARS-CoV-2 cases, but only including data up to the final day of data received in real-time. This was done to evaluate the impact of data revisions on real-time estimates.}

\subsection{Forecasting model}

For the forecasting model, we assume that the time-dependent reproduction number, $R_t$, follows a zero-mean Gaussian process in log-transformed space:
\begin{equation} \label{eq:Rt}
    \ln(R_t) \sim GP\left(0, k(t,t')\right)
\end{equation}
We use the Matern 5/2 covariance function defined by:
\begin{equation}
k(t,t') =  \frac{s_0^2 2^{-3/2}}{\Gamma(5/2)} \left( \frac{\sqrt{2\nu}|t-t'|}{l} \right)^{5/2} K_{5/2}\left( \frac{\sqrt{5}|t-t'|}{l} \right)
\end{equation}
where $s_0$ is the signal standard deviation, $l$ is the correlation time scale, and $K_\nu$ is the modified Bessel function of the second kind and order $\nu$. To implement the Gaussian process model, the value of $\ln(R_t)$ is drawn conditional on the known values of $\ln(R_s)$ for $s=0,\ldots,t-1$ using the method described in \cite{rasmussen2006gaussian}, with Gaussian observation noise with standard deviation $s_n$.

We assume the number of new daily infections $I_t$ on day $t$ follows the Poisson renewal process:
\begin{equation}
I_t \sim \mathrm{Poiss}\left( R_t \sum_{s=1}^{m_g} g_s I_{t-s} \right) \label{eq:It}
\end{equation}
where $g_s$ is the probability mass function for the generation interval ($s=1,\ldots,m_g$) and $m_g$ is the maximum generation interval.

The expected number of infections with a report date on day $t$ ($Z_t$) and the expected number of infections with a hospital admission date on day $t$ ($H_t$) are calculated via forward convolutions:
\begin{align}
Z_t &= \sum_{s=0}^{m_r} u_s I_{t-s}  \label{eq:Zt} \\
H_t &= \sum_{s=0}^{m_h} v_s I_{t-s}  \label{eq:Ht}
\end{align}
where $u_s$ and $v_s$ are the probability mass functions of the distribution of time from infection to reporting and time from infection to hospital admission respectively, and $m_r$ and $m_h$ are their maximum values. 

We assume that the case-hospitalisation ratio $P_t$ on day $t$ follows a Gaussian random walk in log-transformed space with fixed variance $\sigma_P^2$:
\begin{equation}
\ln(P_{t+1}) \sim N( \ln(P_t), \sigma_P^2) \label{eq:Pt}
\end{equation}
{\hl Although this assumption effectively restricts how rapidly the case-hospitalisation ratio can change, we found it to be reasonable based on previous data on SARS-CoV-2 cases and hospitalisations for the past 12 months (noting that this excluded periods in which there were major changes in testing programmes, age-dependent contact rates or vaccination rates, which could drive a more rapid change in $P_t$). We also monitored the estimated value of $P_t$ and associated model fit throughout the season (see Results). }

We assume that the observed number of cases, $C_t$, and new hospital admissions, $A_t$, on day $t$ are drawn from conditionally independent negative binomial distributions (parameterised by mean and dispersion):
\begin{align}
C_t &\sim  \mathrm{NegBin}\left( p_c \omega_{c,t} Z_t, k_c \right), \label{eq:Ct} \\
A_t &\sim  \mathrm{NegBin}\left( p_c \omega_{h,t} P_t H_t, k_h \right), \label{eq:At}
\end{align}
where $p_c$ is the proportion of infections that are reported as cases (assumed to be fixed), and $\omega_{c,t}$ and $\omega_{h,t}$ are day-of-the-week effects for reported cases and hospital admissions, respectively. Note that the latent variables $Z_t$ and $H_t$ include all infections, whereas the observed variables $C_t$ and $A_t$ represent the subset of infections that were reported as cases or hospitalised, respectively. The case ascertainment ratio $p_c$ is non-identifiable from the data, and so we set $p_c=1$. {\hl This means that the latent variable $I_t$ should be interpreted as the daily incidence of new infections that will eventually be reported as cases, as opposed to the actual number of infections, which is unknown. In reality, case ascertainment could vary over time, which could affect forecast performance. However, as surveillance effort over the season is consistent, we believe variations in case ascertainment are likely to be relatively small and to occur slowly relative to the time scale for a seasonal epidemic and the forecasting time horizon of 28 days. }

Eqs. \eqref{eq:Ct}--\eqref{eq:At} define a likelihood function for a given observed value of $C_t$ in terms of $\omega_{c,t} Z_t$ and for a given observed value of $A_t$ in terms of $\omega_{h,t} P_t H_t$. Note that the model includes the flexibility to specify either or both of these distributions as a Poisson distribution by setting the relevant dispersion parameter $k=\infty$. 

We estimate the day-of-the-week effects, $\omega_{c,t}$ and $\omega_{h,t}$, directly from the data via
\begin{equation}
    \omega_{*,t} = \frac{7 \sum_{s=t\ \mathrm{mod} \ 7} \tilde{X}_s }{\sum_t \tilde{X}_t}
\end{equation}
where $\tilde{X}_t$ is the relevant observed quantity (either reported cases for  $\omega_{c,t}$ or hospital admissions for $\omega_{h,t}$) on day $t$. The summations are taken over an integer number of weeks leading up to the origin date, up to a maximum of 16 weeks depending on how much past data is available. This assumes that the day-of-the-week effects are constant over this time period, which may not always be the case. {\hl One alternative would be to fit a time-varying day-of-the-week effect model. However, we decided not to do this as it would introduce substantial additional complexity and would risk masking real epidemic trends by subsuming them into fitted day-of-the-week effects. Another alternative would be to ignore day-of-the-week effects completely, or aggregate data to weekly periods. However, this would lose important information and compromise the model's ability to pick up on short-term changes in the epidemic trend.}

For details of the initialisation and fitting method, see Supplementary Material Sec. S1. Parameter values used in the model are shown in Table 1. {\hl The parameters for the generation interval and delay distributions were estimated from the literature (see Supplementary Material Sec. 1.3 for details). The parameters of the Gaussian process $s_0$ and $l$, and the observation noise parameters $k_c$ and $k_h$ were calibrated manually to achieve a good visual model fit. In principle, these parameters could be inferred from the data with suitable uncertainty. However, we chose not to do this due to the need to balance model complexity with computational limitations for real-time deployment. The parameter $P_{CV}$ is for initialisation purposes only and does not affect model fit or forecasts, provided that the model is run for a sufficient time period. }

\begin{table}
\scriptsize
\begin{tabular}{llll}
\hline
{\bf Pathogen-specific parameters} & {\bf SARS-CoV-2} & {\bf Influenza} & {\bf RSV} \\
\hline
Generation interval & $3.3 \pm 3.5$ days & $2.6 \pm 1.3$ days & $7.5 \pm 2.1$ days\\
Time from infection to reporting & $6.3 \pm 3.1$ days & -  & -  \\
Time from infection to admission & $6.3 \pm 3.7$ days & $3.6\pm 2.1$ days & $6.9\pm 2.7 $ days \\
\hline
{\bf Pathogen-independent parameters} &&& \\
Maximum generation interval & \multicolumn{3}{c}{ $m_g=15$ days } \\
Maximum time from infection to reporting & \multicolumn{3}{c}{$m_r=25$ days}  \\
Maximum time from infection to admission & \multicolumn{3}{c}{$m_h=25$ days } \\
Gaussian process signal s.d. & \multicolumn{3}{c}{$s_0=0.1$} \\
Gaussian process autocorrelation time scale & \multicolumn{3}{c}{$l=30$ days} \\
Gaussian process observation noise s.d. & \multicolumn{3}{c}{$s_n=0.001$} \\
Probability of infection being reported & \multicolumn{3}{c}{$p_c=1$} \\
Dispersion parameter for observed daily cases & \multicolumn{3}{c}{$k_c=25$} \\
Dispersion parameter for observed daily admissions & \multicolumn{3}{c}{$k_h=25$} \\
Coefficient of variation of initial case-hospitalisation ratio & \multicolumn{3}{c}{$P_{CV}=0.025$}  \\
\hline
\end{tabular}
\caption{Table of forecasting model parameter values. Parameters shown as $\mu \pm \sigma$ show the mean ($\mu$) and standard deviation ($\sigma$) of the assumed distribution.  Note the model nominally assumes $p_c=1$ (meaning that all modelled infections are eventually reported as cases), but this parameter does not affect model output for observed variables (daily cases and hospital admissions), only for the latent daily infection variable $I_t$, which we treat as being identifiable only up to multiplication by an unknown case ascertainment ratio, which we assume to be constant.}
\end{table}

We evaluated probabilistic forecasts $y$ against subsequently observed data $y^\mathrm{obs}$ using the continuous ranked probability score (CRPS), defined as
\begin{equation}
\mathrm{CRPS}\left(y, y^\mathrm{obs} \right) = \int_{-\infty}^\infty \left(F(y) - \mathbb{I}\left(y>y^\mathrm{obs}\right)\right)^2 dy
\end{equation}
where $F(y)$ is the cumulative distribution function for the forecast and $\mathbb{I}(.)$ denotes the indicator function. CRPS was calculated numerically using $2000$ randomly selected particles. The CRPS was calculated on log-transformed values so that it can be interpreted a measure of relative rather than absolute error in forecast counts \cite{bosse2023scoring}.

The code used to produce the results in this article is publicly available \cite{github-repo}.

\section{Results}

We begin by summarising the overall epidemic trends of SARS-CoV-2, influenza and RSV as seen retrospectively in the full season's data (Sec. \ref{sec:overall_trends}). Then we describe the real-time weekly reporting that we carried out throughout the season (Sec. \ref{sec:real_time_reporting}, and evaluate the performance of trend estimates and forecasts made in real-time against subsequently available data (Sec. \ref{sec:trend_evaluation}--\ref{sec:forecast_evaluation}).

\subsection{Overall epidemic trends} \label{sec:overall_trends}

\begin{figure}
    \centering
    \includegraphics[width=\linewidth]{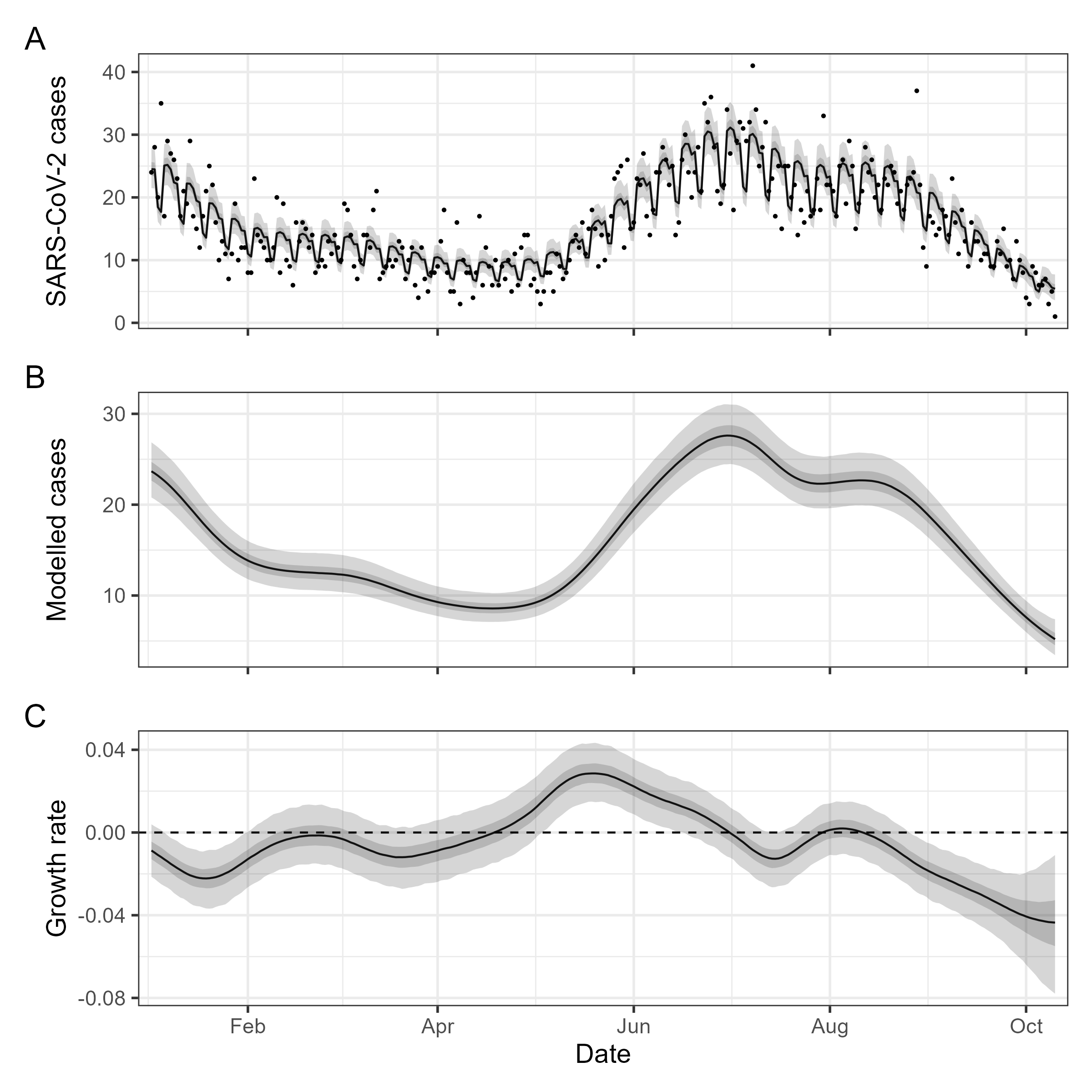}
    \caption{Modelled trends in SARS-CoV-2 cases fitted to full season data. (A) Daily SARS-CoV-2 cases (points) and modelled trends {\hl (i.e. the expected value of the time-series)} in SARS-CoV-2 cases using a Bayesian P-spline model, including a day-of-the-week effect (line and shaded region). (B) Modelled trend in SARS-CoV-2 cases (with day-of-the-week effect removed yet still included in the model). (C) Growth rate inferred from the modelled trend in SARS-CoV-2 cases (with day-of-the-week effect removed). The dashed line indicates a growth rate of zero reflecting the threshold between epidemic growth or decline. For all panels we include the model posterior distribution's median (line) and 50\% and 95\% credible intervals (dark shaded and light shaded regions).}
    \label{fig:case-trends-overview}
\end{figure}

\begin{figure}
    \centering
    \includegraphics[width=0.8\linewidth]{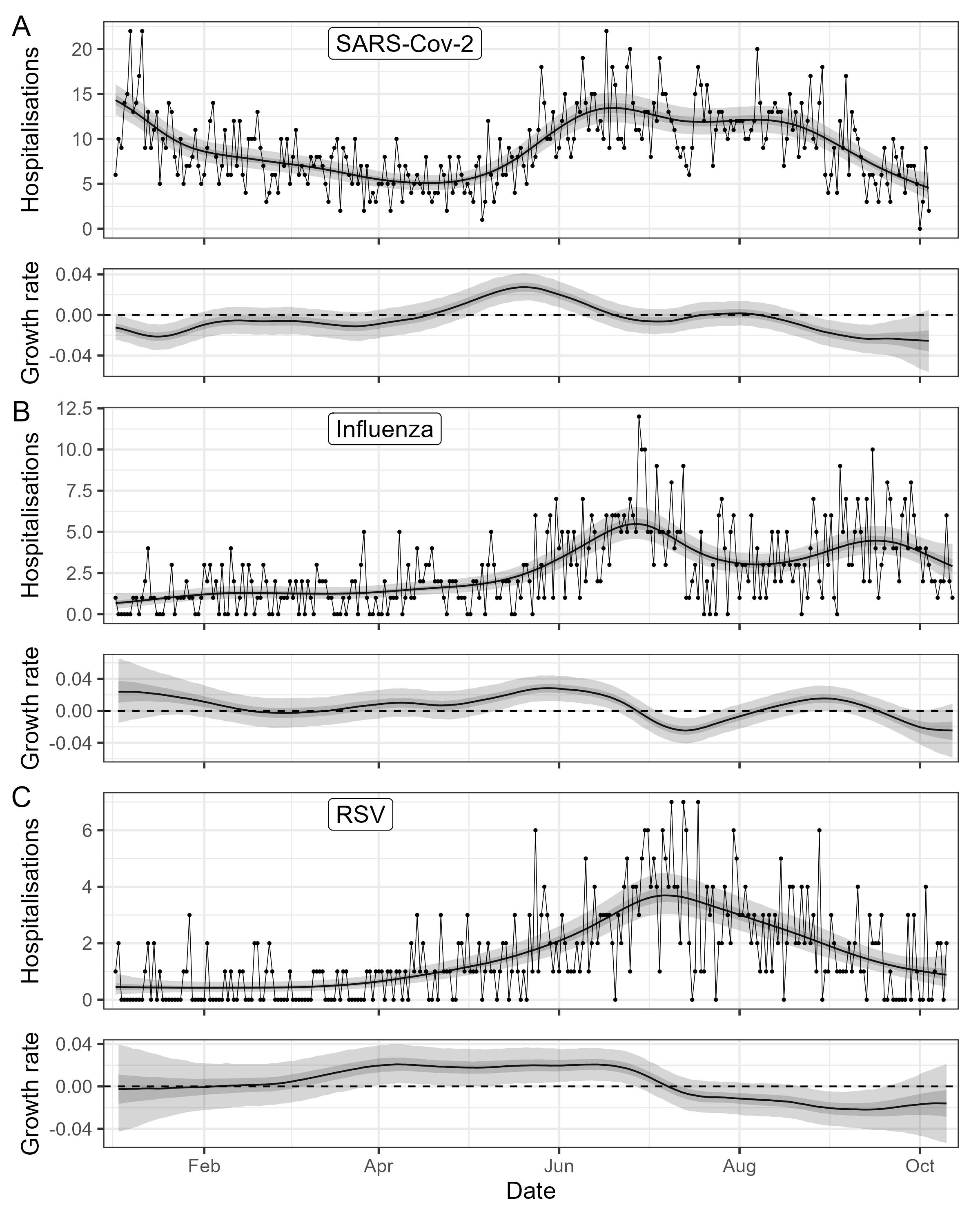}
    \caption{Modelled trends in hospitalisations fitted to full season data for: (A) SARS-CoV-2; (B) influenza; and (C) RSV. In the top panel for each pathogen we plot the daily number of hospitalisations (points) and modelled trend {\hl (i.e. the expected value of the time-series)} in hospitalisations (line and shaded regions) using a Bayesian P-spline model that does not include a day-of-the-week effect. In the bottom panel for each pathogen we plot the growth rate inferred from the modelled trend in hospitalisations. The dashed line indicates a growth rate of zero (the threshold between epidemic growth or decline). For all panels we include the model posterior distribution's median (line) and 50\% and 95\% credible intervals (dark shaded and light shaded regions).}
    \label{fig:hosp-trends-overview}
\end{figure}

SARS-CoV-2 cases began increasing in New Zealand from April 2025 (Figure \ref{fig:case-trends-overview}). Cases increased until late-June/early-July when a peak in modelled cases {\hl (i.e. the expected number of daily cases excluding day-of-the-week effects)} was reached (approximately 28 cases per day). Cases decreased for approximately two weeks before stabilising, following a small resurgence in the epidemic growth rate from mid-July. SARS-CoV-2 cases were then stable until mid-August, before declining until the end of the study period (October). 
SARS-CoV-2 hospitalisations exhibited broadly similar temporal dynamics to SARS-CoV-2 cases (Figure \ref{fig:hosp-trends-overview}). The rise in SARS-CoV-2 case and hospitalisations coincided with an increase in the proportion of genomically sequenced cases that were due to the NB.1.8.1 lineage, which increased from low levels in early May to around 90\% by the end of July \cite{phf_covid_genomics_report62}. This introduction of a new variant with a transmission advantage over established variants potentially contributed to overall growth in SARS-CoV-2 epidemic activity.

There were two peaks in influenza hospitalisations over the course of the season: one in late-June; and one in mid-September. There was only one peak in RSV hospitalisations, occurring in early-July. It is not clear when the modelled daily number of influenza and RSV hospitalisations began increasing {\hl(i.e. growth rate greater than 0)} due to the low daily numbers of outcomes in these time series and resulting high level of uncertainty estimated by the statistical smoothing model {\hl for the time when the growth rate became positive}. {\hl Modelled trends in influenza and RSV hospitalisations were sensitive to the assumed prior distribution for the parameter controlling the smoothness of the second-order random-walk, $\tau$. When the mean of the prior distribution for $\tau$ was greater (i.e., less smooth), the model estimated sharper changes in the growth rate and wider credible intervals (see Supplementary Figure S1).}

\subsection{Real-time reporting} \label{sec:real_time_reporting}

We analysed case and hospitalisation time series in real-time during the 2025 season, producing 15 reports for distribution to the New Zealand Ministry of Health and Te Whatu Ora Health New Zealand from 6 June (Report 1) to 3 October 2025 (Report 15). Each report presented estimates of past and current trends in cases and hospitalisations --- including estimates of growth rates, doubling/halving times, and the probability that the growth rate was greater than zero (cases/hospitalisations growing) --- and, where available, 28-day ahead forecasts. An example report (Report 6, produced on 1 August 2025) is included in Supplementary Material Sec. S3. We chose Report 6 because earlier reports did not include forecasts as these were {\hl being developed and tested on previous season's data}. Due to reporting delays, the results in this report were based on data up to 27 July 2025 for SARS-CoV-2 and 19 July 2025 for influenza and RSV. Forecasts show a 28-day time horizon from these respective origin dates.

\subsection{Evaluation of real-time estimates of past and current trends} \label{sec:trend_evaluation}

\begin{figure}
    \centering
    \includegraphics[width=0.75\linewidth]{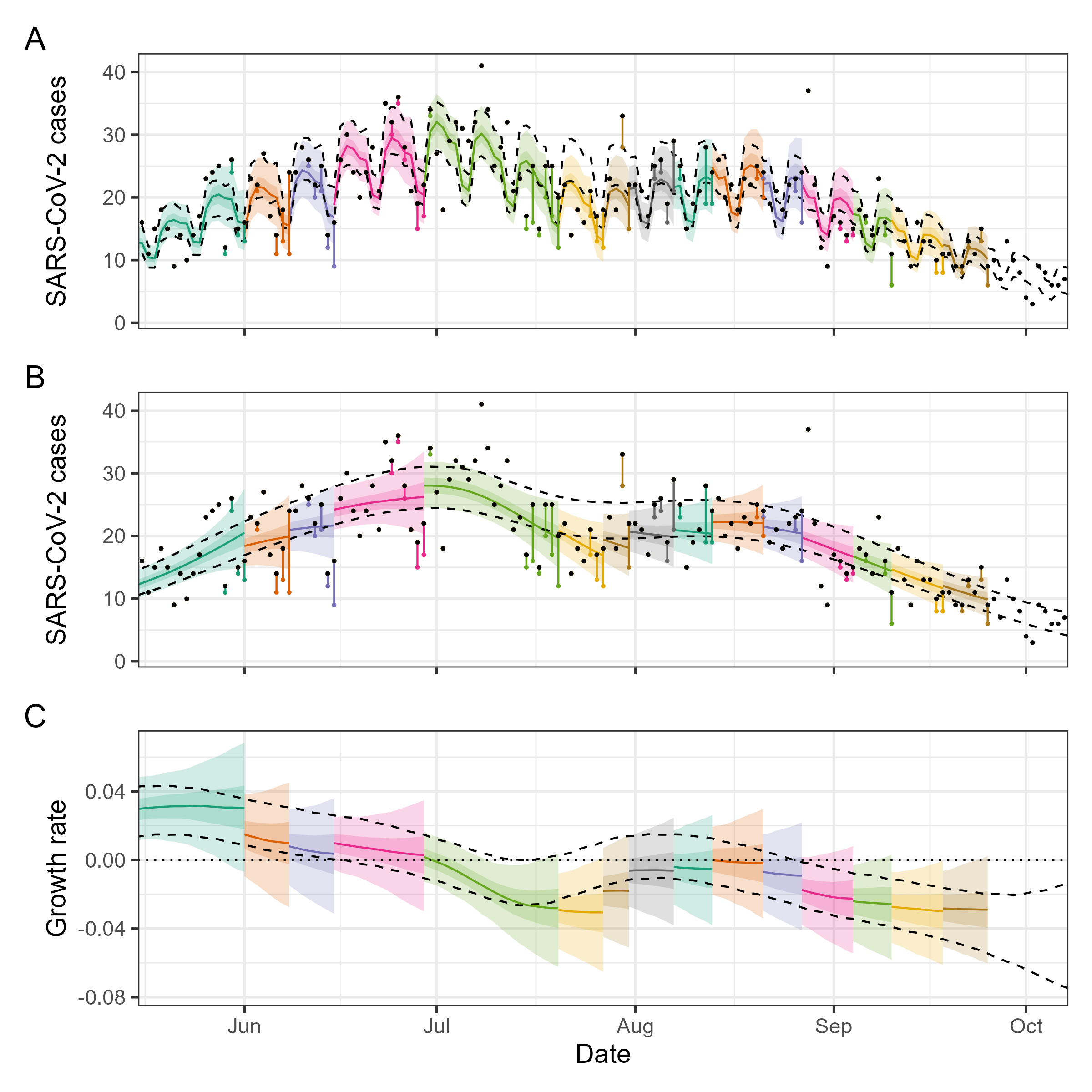}
    \caption{Modelled trends in SARS-CoV-2 cases reported in real-time over the course of the season. (A) Modelled trends in SARS-CoV-2 cases including day-of-the-week effect (lines and shaded region). (B) Modelled trends in SARS-CoV-2 cases with day-of-the-week effect removed (lines and shaded region). (C) Growth rate (lines and shaded region) inferred from the modelled trend in SARS-CoV-2 cases (with day-of-the-week effect removed). For all panels real-time estimates (coloured lines and shaded regions) are compared to 95\% credible intervals of end-of-season estimates (black dashed lines). For all real-time estimates (coloured by round), we include the model posterior distribution's median (line) and 50\% and 95\% credible intervals (dark shaded and light shaded regions). We include the real-time model estimates up to the final day of data (for periods greater than the previous round's final day of data). In (A) and (B) the daily SARS-CoV-2 cases for the final end-of-season dataset (black points) and the daily SARS-CoV-2 cases in the real-time datasets (points coloured by round) are connected by lines to highlight revisions to data used in real-time analyses. For example, during the fourth round (first pink) there are visible pink data points connected directly to the black end-of-season data; the data in pink is the data available when models were estimated, the data was revised upwards in the following round where it matched the end-of-season data. Note that when coloured points (and connecting lines) can not be seen then there has been no revisions to the data. When a coloured point can be seen and is the same colour as the real-time estimate for the same period (and connected directly to a black data point), it indicates that the number of cases for a specific day was revised (upwards in most instances) in the following round of data, but was not revised in any further rounds of data).
    }
    \label{fig:case-trends-real-time}
\end{figure}

\begin{figure}
    \centering
    \includegraphics[width=0.6\linewidth]{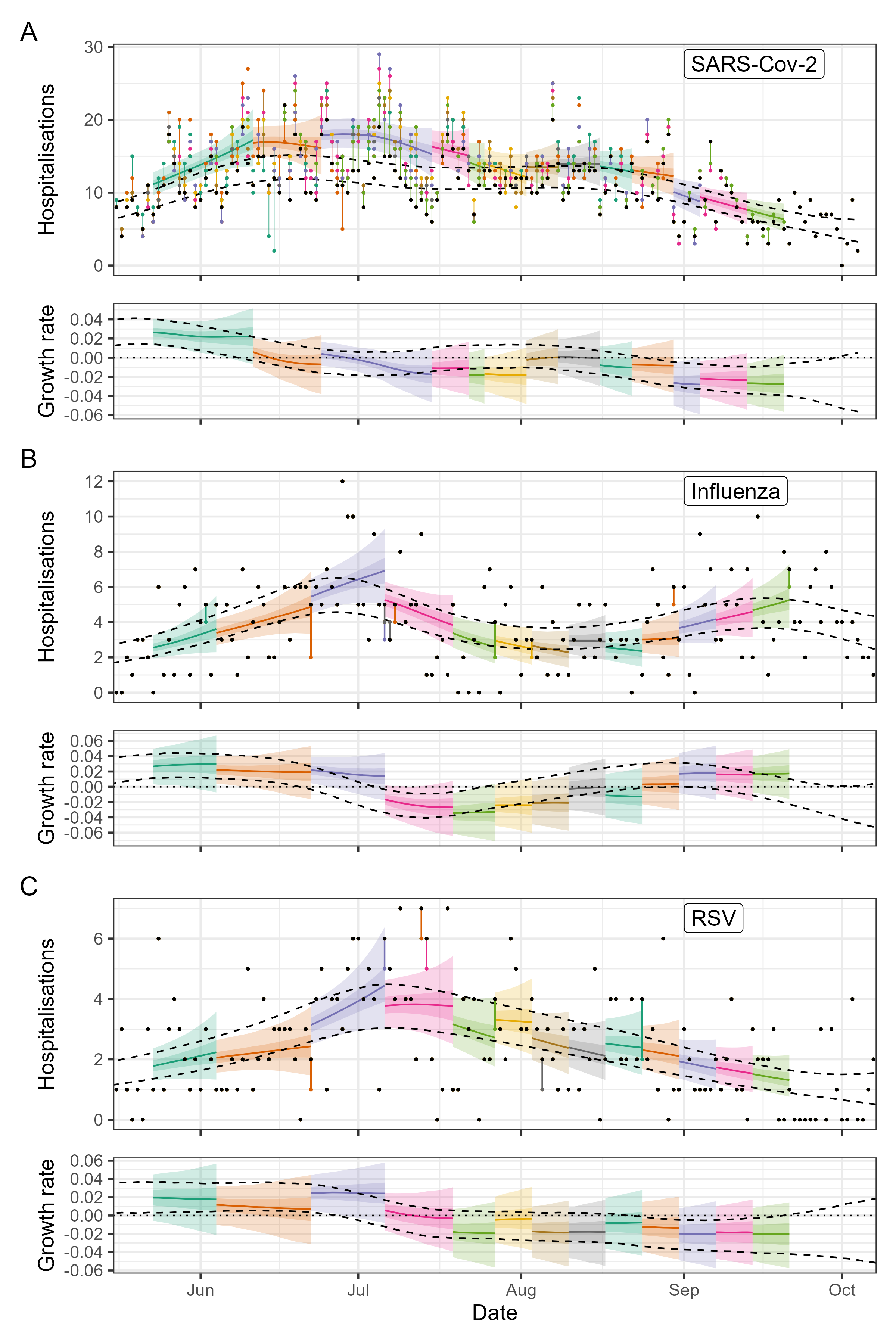}
    \caption{Modelled trends in hospitalisations reported in real-time over the course of the season. Modelled trends in hospitalisations are shown for: (A) SARS-CoV-2; (B) influenza; and (C) RSV. For each pathogen we plot: (top panels) modelled trends in hospitalisations (lines and shaded region); and (bottom panels) growth rate (lines and shaded region) inferred from the modelled trend in hospitalisations. For all panels real-time estimates (coloured lines and shaded regions) are compared to 95\% credible intervals of end-of-season estimates (black dashed lines). For all real-time estimates (coloured by round) we include the model posterior distribution's median (line) and 50\% and 95\% credible intervals (dark shaded and light shaded regions). We include the real-time model estimates up to the final day of data (for periods greater than the previous round's final day of data). The daily number of hospitalisations for the final end-of-season dataset (black points) and the daily number of hospitalisations in the real-time datasets (points coloured by round) are connected by lines to highlight revisions to data used in real-time analyses. Note that when coloured points (and connecting lines) can not be seen then there has been no revisions to the data. Note that for SARS-CoV-2 hospitalisations some data were revised in multiple rounds and so many points can be seen for the same time point.}
    \label{fig:hosp-trends-real-time}
\end{figure}

We compared estimates made in real-time over the course of the season to final end-of-season estimates made with data available after season reporting had finished (final data received 23 October 2025). Real-time estimates of trends in SARS-CoV-2 cases showed good agreement to the end-of-season estimate, with overlapping 95\% credible intervals (Figure \ref{fig:case-trends-real-time}). The largest disagreements between real-time and end-of-season estimates were observed in late-July (Reports 5--7), when real-time estimates suggested epidemic activity was decreasing faster than estimated at the end of the season. During this period there was a transient increase in the epidemic growth rate. There were also substantial revisions to the data received in real-time (counts were revised upwards in the following week's data) (Figure \ref{fig:case-trends-real-time} and Supplementary Figure S3), which biased our real-time estimates downwards {\hl (see Supplementary Figure S2)}. 

Real-time estimates of expected SARS-CoV-2 hospitalisations were higher than the end of season estimate for most rounds of reporting (Figure \ref{fig:hosp-trends-real-time}). This was likely due to significant revisions to data (Supplementary Figure S4) made over the entire study period (e.g., data available in the first round of reporting was revised multiple times over the study period). RSV and influenza hospitalisation data was far more stable, with fewer revisions made to the data (and data for a single date revised at most once). Accordingly, real-time estimates of expected hospitalisations and their trends for influenza and RSV showed greater agreement with end-of-season estimates (compared to SARS-CoV-2). The greatest disagreement was observed close to the peaks in hospitalisations, where real-time estimates suggested epidemic activity was continuing to increase. However, the lower credible intervals of these estimates were   still captured within the end-of-season estimates.

\subsection{Evaluation of real-time 28-day ahead forecasts} \label{sec:forecast_evaluation}

\begin{figure}
    \centering
    \includegraphics[width=1.05\linewidth]{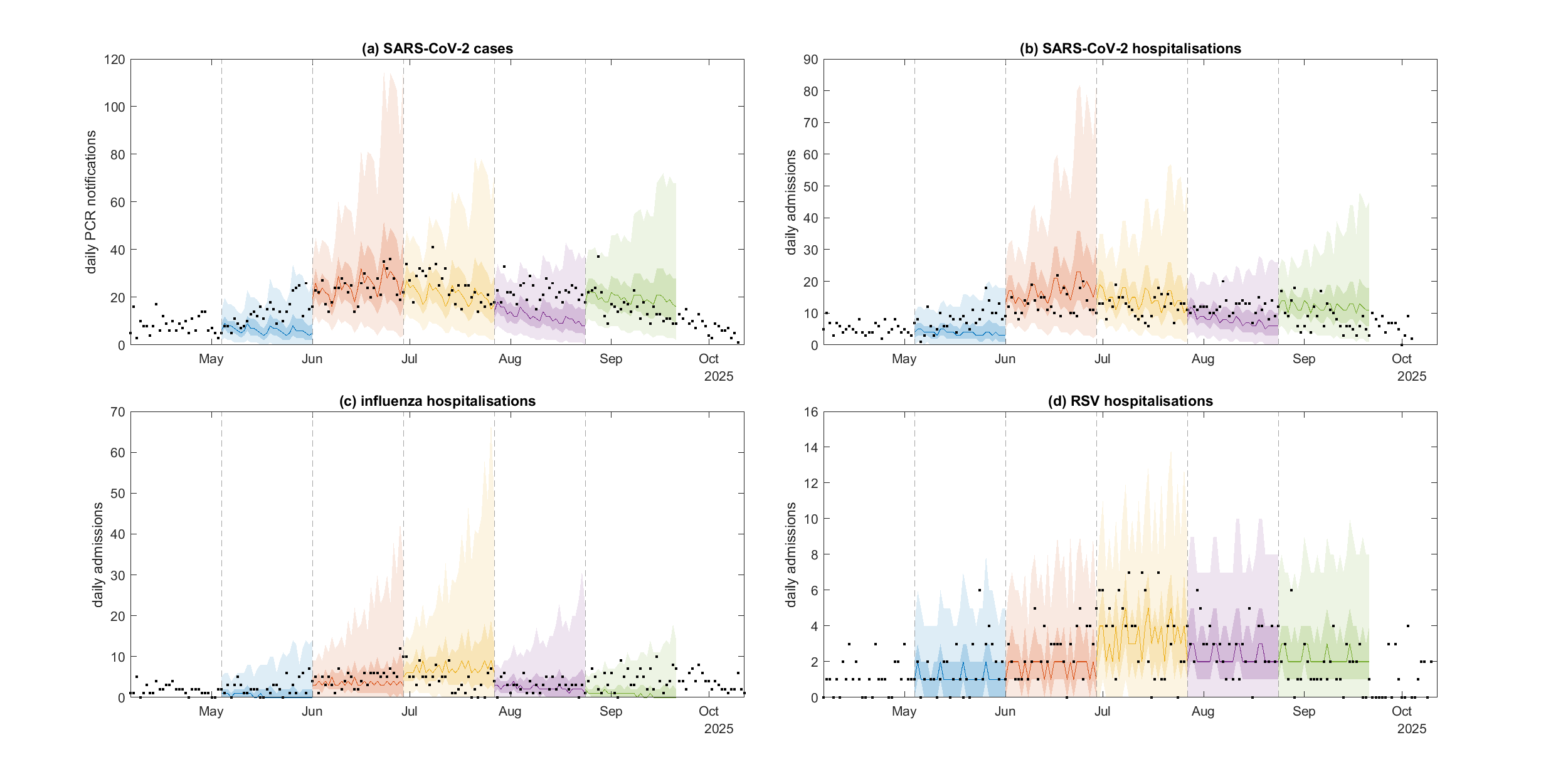}
    \caption{In-season 28-day ahead forecasts (coloured lines and bands) alongside subsequently observed data (black points) for: (a) daily SARS-CoV-2 cases; (b) daily SARS-CoV-2 hospitalisations; (c) daily influenza hospitalisations; (d) daily RSV hospitalisations. Each coloured block shows a forecast generated at a specific origin date (shown by the vertical dashed lines) showing the posterior median (line) and 50\% and 95\% credible intervals (dark shaded and light shaded regions). }
    \label{fig:forecasts_all_pathogens}
\end{figure}

{\hl To evaluate forecast performance, we compared 28-day ahead forecasts produced for weekly origin dates (20 rounds from 4 May 2025 to 14 September 2025) against subsequently available data for the whole season (Figure \ref{fig:forecasts_all_pathogens}). For ease of visualisation, Fig.  \ref{fig:forecasts_all_pathogens} shows every fourth round so that the forecast periods do not overlap in time; Supplementary Figures S6--S9 show forecasts for all rounds. }

At the beginning of the season, SARS-CoV-2 forecasts were slow to pick up the increase in epidemic activity, and the early forecasts (e.g. blue blocks in Fig. \ref{fig:forecasts_all_pathogens}a-b) underestimated subsequent data, although the data mostly fell within the 95\% credible interval. For the remainder of the season, forecasts captured the data reasonably well, with the exception of a few rounds where the forecast systematically underestimated or overestimated the data {\hl (e.g. purple block in Fig. \ref{fig:forecasts_all_pathogens}a)}. {\hl However, overall interval coverage properties were good, with 99.5\% of data points for cases and 97.7\% for hospitalisations falling within the 28-day ahead forecast 95\% credible intervals (see Supplementary Table S1). An example set of model outputs for SARS-CoV-2 including latent variables is shown in Supplementary Figure S5. This shows that the the model provided a good fit to training data and the estimated case-hospitalisation ratio $P_t$ varied little over the relevant time period.  }
 
Influenza forecasts {\hl (Figure \ref{fig:forecasts_all_pathogens}c)} performed reasonably well, with good interval coverage throughout the season (99.6\% of data fell within the 28-day ahead 95\% credible intervals -- see Supplementary Table S1), although with wider credible intervals (compared to SARS-CoV-2). This is likely a consequence of the relatively small numbers of influenza hospitalisations in the data {\hl (recalling that influenza and RSV data were obtained from a regional sentinel surveillance programme, and so are not comparable with SARS-CoV-2 data)}. For the 29 June origin date (yellow block in Figure \ref{fig:forecasts_all_pathogens}c), which was around the peak in influenza activity, the forecast exhibited very wide credible intervals and somewhat overestimated subsequent data. The following week's forecast (yellow block in Supplementary Figure S8b) provided a better prediction of the subsequent trend.  {\hl Previous analyses have found that respiratory disease forecasts tend to perform worse and/or have high uncertainty near to the epidemic peak \cite{lopez2024challenges}. However, although this was true of some of our forecasts, it was not universally true of all forecasts near the peak (see Supplementary Figures S6--S9), potentially because these peaks were quite gradual with relatively modest rates of growth and decay and high levels of noise in the data. }

{\hl RSV forecasts (Figure \ref{fig:forecasts_all_pathogens}d) also performed well at predicting the trend and displayed good interval coverage (98.2\% of data fell within the 28-day ahead 95\% credible intervals -- see Supplementary Table S1). The 95\% credible forecast intervals almost always included zero, reflecting the high frequency of zero counts throughout the relevant time period. RSV forecasts tended to be less sensitive to short-term fluctuations in the data compared  to SARS-CoV-2 and influenza, likely due in part to its longer generation interval. }

\begin{figure}
    \centering
    \includegraphics[width=\linewidth]{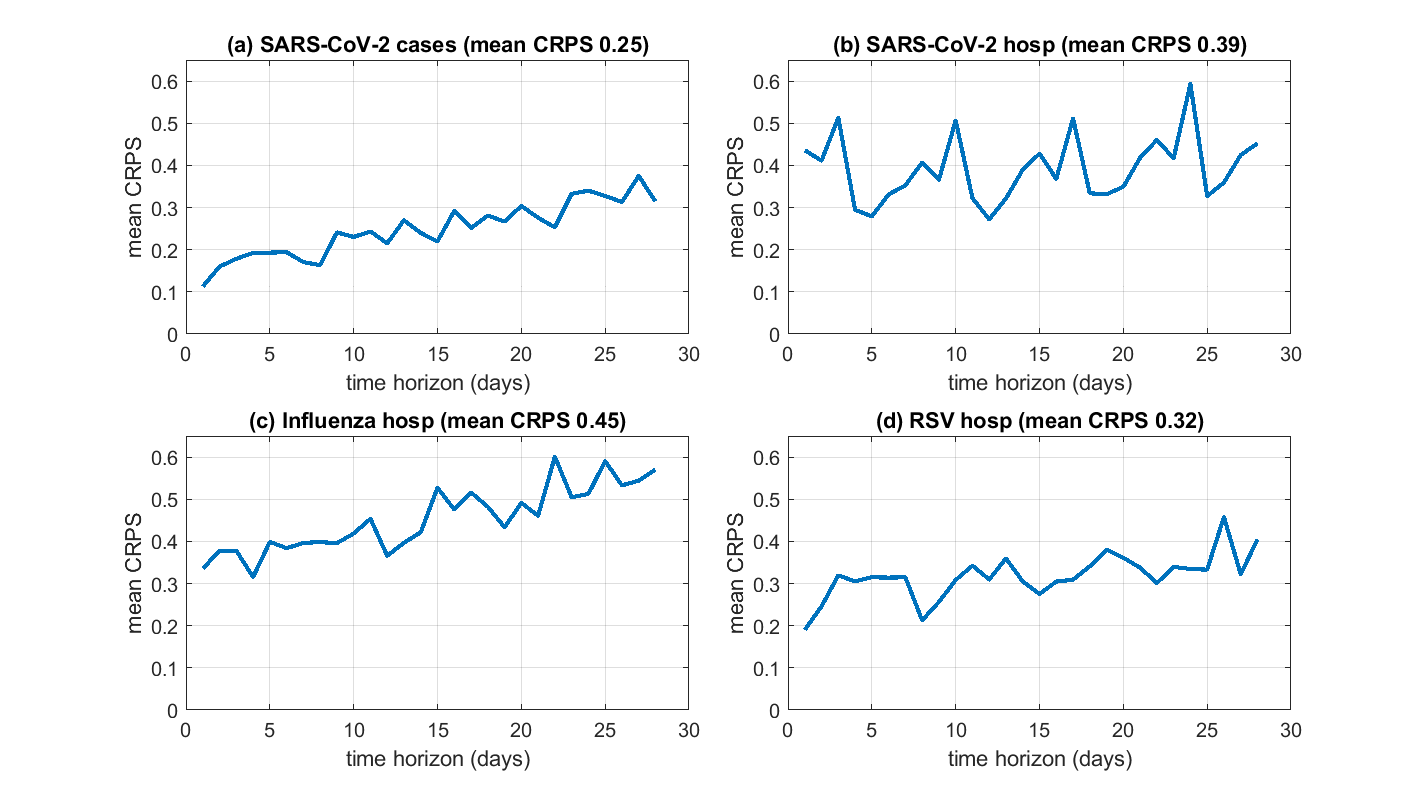}
    \caption{Mean continuous ranked probability scores (CRPS) at different forecast horizons for each of the forecast targets: (a) SARS-CoV-2 cases; (b) SARS-CoV-2 hospitalisations; (c) influenza hospitalisations; (d) RSV hospitalisations. CRPS values were calculated on log-transformed data and averaged over all rounds for a given forecast horizon. The value for mean CRPS shown in the text above each panel show the CRPS averaged over all rounds and forecast horizons for that target. }
    \label{fig:scores}
\end{figure}

Figure \ref{fig:scores} shows the mean CRPS for each forecast target as a function of the forecast time horizon, averaged over all rounds. The CRPS can be interpreted as a measure of the relative difference between the probabilistic forecast distribution and the subsequent data point, i.e. smaller CRPS indicates better forecast performance. For most forecast targets there was an increasing trend of mean CRPS with time horizon, indicating that as expected, forecasts tended to perform worse as the time horizon increased. {\hl Overall, the forecasts for SARS-CoV-2 cases (Figure \ref{fig:scores}a) cases had the best CRPS (mean 0.25). The forecast for SARS-CoV-2 hospitalisations (Figure \ref{fig:scores}b) had a higher CRPS (mean 0.39) and showed signs of weekly periodicity with spikes at time horizons of 3, 10, 17 and 24 days. This could be due to a day-of-the-week effect that was not fully captured by the model, or due to historical revision of data primarily occurring on the same day of the week. Influenza had the worst CRPS of the four targets (mean 0.45), while RSV was intermediate (mean 0.32). The fact that forecast coverage was somewhat higher than the nominal interval coverage at the 50\% and 95\% level for all four targets (see Supplementary Table S1) suggests that the forecast model is overly conservative in its uncertainty quantification, which will contribute to poorer CRPS scores. }

\section{Discussion}

We have presented the methods and results of a 2025 New Zealand winter situational assessment programme for SARS-CoV-2, influenza and RSV. This was the first year of the programme; prior to 2025, New Zealand lacked capacity for real-time model-based situational assessment and forecasting for seasonal respiratory diseases. This programme complements traditional surveillance of respiratory disease \cite{phf2025acute} by improving our ability to interpret and predict epidemiological trends in real-time.   

Our forecasts and estimates of current trends performed reasonably well overall, but were worst during periods of rapid change in the epidemic growth rate. This was not a surprising finding. Both models used in our analyses are fundamentally designed to estimate the current trend in the data; the forecast model assumes that this trend continues into the near future, with some gradual reversion to a prior distribution for the effective reproduction number over time. As a consequence, our models cannot predict trends before they start to become apparent in the data. {\hl While the two models had different objectives, and are not directly comparable, they were often in broad agreement in the direction epidemic trends (i.e. growing or declining). Small disagreements were observed between the models for SARS-CoV-2 cases in early-June; central estimates of the growth rate were greater than zero, but central estimates of the forecasts predicted decreases in cases. However, the uncertainty in both model's estimates led to substantially overlapping credible intervals, resulting in a relatively consistent epidemiological picture or assessment.} 

{\hl Estimates of current trends in SARS-CoV-2, influenza and RSV made by the trend analysis model during 2025 will have been influenced by trends in SARS-CoV-2 data from previous seasons. The model used for trend analysis penalises changes in the epidemic growth rate, with the degree of penalisation set by a parameter ($\tau$). As such, estimates of current trends are sensitive to the value of $\tau$. The models fit to SARS-CoV-2 cases and hospitalisations estimated $\tau$ from the previous three years of SARS-CoV-2 data and therefore estimates will have been influenced by the epidemic dynamics (or the degree to which the growth rate varies over time) over this time period. For influenza and RSV hospitalisations, we only had access to less than one year of data, and therefore assumed an informative prior for $\tau$ (and the overdispersion in the negative binomial distribution) to achieve model convergence. These informative prior distributions were based on posterior distributions from the model fit to SARS-CoV-2 hospitalisations, which may not be indicative of the expected dynamics for influenza and RSV. In future years, as more data for influenza and RSV hospitalisations becomes available, we will aim to estimate all parameters for each pathogen individually.}

Our models are likely to be biased when substantial revisions occur to data received in real-time relative to the final data. SARS-CoV-2 cases recorded in the most recent week of data were regularly revised upwards in the following round of data (and influenza and RSV hospitalisations to a lesser extent). This likely biased estimates of current trends and forecasts downwards, but estimates of past trends would have been less affected. In contrast, SARS-CoV-2 hospitalisations were revised frequently, with the same day of data often revised over multiple rounds, biasing estimates of past, current and future trends.  Statistical methods could be developed to quantify revisions in hospitalisation or case counts over time. {\hl Existing methods rely on having sufficient data about when and how much counts are revised (i.e. two date variables, one for the date of occurrence and for the date of receipt). Our methods could correct for the expected degree of revision in future seasons based on the 15 rounds of data received in 2025. For example, this could be done by fitting a model for the probability $P(n,m,s)$ that the final count for a given date $t$ is $n$ given that the observed count is $m$ according to data available at time $t+s$. } This would enable these changes and associated uncertainty to be accounted for when estimating epidemic trends in real-time, as has been done elsewhere \cite{johnson2025baseline}. {\hl However, we did not have sufficient information to perform such an analysis in advance of the 2025 season because we only had access to a single dataset for previous seasons, with no information on how those data would have been received and revised in real-time. }

The timing of peaks in epidemic activity of respiratory pathogens may be affected by school holidays, {\hl which have been estimated to reduce transmission of seasonal respiratory diseases \cite{jackson2016relationship}}. School holidays in New Zealand ran from 28 June to 13 July 2025 and peaks in SARS-CoV-2 cases and influenza and RSV hospitalisations occurred around this period. School holidays can affect both transmission (by altering contact patterns), as well as testing and healthcare-seeking behaviour, which will influence case and hospitalisation time-series. Without more detailed data on behaviour \cite{Eales2025-bt} or underlying infection levels (as opposed to cases) \cite{Eales2024-gv}, we cannot disentangle the underlying reasons for these changes. It is also possible that similarly-timed peaks would have occurred irrespective of school holidays due to depletion of the susceptible population. 

In 2025, an extended influenza season was experienced in southern hemisphere countries, including New Zealand, driven by the emergence of the antigenically distinct subclade K of influenza A/H3N2. Genomic data indicates that this virus was likely seeded into New Zealand from Australia in August 2025 \cite{dapat2025extended}, and caused influenza activity to continue into November and December at unusually high levels for the time of year. At the same time, northern hemisphere countries experienced unusually early influenza activity as a result of the same variant \cite{hay2025evaluation,Fieldhouse2025-dp}. Our results showed an increase in influenza activity in the last few weeks of our analysis (early September). However, most of the subclade K-driven influenza activity in New Zealand occurred after our regular analysis and reporting had ended for the season. 

In future seasons, we aim to build on the platform established by this work. Objectives for future work include: expanding the hub approach to include more forecasting models and reporting an ensemble forecast; developing statistical methods to account for revisions to data received in real-time; expanding the types of analysis reported, for example to {\hl include the size and timing of the epidemic peak as explicit forecast targets}; and incorporating national data on influenza hospitalisations from the National Minimum Dataset (NMDS) \cite{nmds_website}. {\hl The ACEFA forecasting hub also runs a winter situational assessment programme for Australian states and territories, providing weekly reports comparable to those described here for New Zealand (e.g. see \cite{Eales2026-bi,henderson2026winter}). We are working towards producing a combined trans-Tasman weekly report covering both countries, subject to data sharing and confidentiality agreements. This would enable timely exchange of surveillance intelligence between Australia and New Zealand and build joint capability in real-time epidemic analytics for public health partners.}

\subsection*{Data availability statement}
The code used to produce the results in this article is publicly available \cite{github-repo}.
The raw data cannot be shared publicly for confidentiality reasons, but can be requested from Te Whatu Ora (Health New Zealand) at: \url{data-enquiries@tewhatuora.govt.nz}.

\subsection*{Conflict of interest statement}
We declare no conflicts of interest.

\subsection*{Ethics statement}

This research was carried out under a data sharing agreement between Te Whatu Ora (Health New Zealand) and ACEFA. The terms of this agreement did not require ethics approval for analysis of anonymised, aggregated data.

\subsection*{Funding statement}
MJP was supported by a grant from the Marsden Fund (24-UOC-020) and from Te Niwha Infectious Diseases Research Platform, co-hosted by PHF Science and the University of Otago and provisioned by the Ministry of Business, Innovation and Employment, New Zealand (TN/P/24/UoC/MP). OE was supported by a University of Melbourne McKenzie fellowship. FMS was supported by the National Health and Medical Research Council of Australia through the Investigator Grant Scheme (Emerging Leader Fellowship, 2021/GNT2010051). This research is supported by the Australia--Aotearoa Consortium of Epidemic Forecasting and Analytics (ACEFA), a National Health and Medical Research Council of Australia Centre of Research Excellence (2035303).

\subsection*{Acknowledgements}
We are grateful to Te Whatu Ora (Health New Zealand), Manat\=u Ora (Ministry of Health), and Public Heath and Forensic Science for providing weekly data updates throughout the 2025 winter respiratory season and feedback on the content of the weekly reports, in particular Andy Anglemeyer, Harriette Carr, Laura Cleary, Susan Jack, Bex Joslin, Tomasz Kiedrzynski, Zoe Kumbaroff, Andrea McNeill, Natasha Rafter, Imogen Roth, Tanisha Selvaratnam and Fiona Wild. The authors are grateful to Rob Hyndman for feedback on the forecasting model methodology and to Alex Kazemi and three anonymous reviewers for comments on an earlier version of this manuscript.

\bibliography{references}

\end{document}


\newpage

\maketitle

\startcontents[sections]
\printcontents[sections]{l}{1}{\setcounter{tocdepth}{2}}

\clearpage

\section{Supplementary methods} \label{sec:supp_methods}

\subsection{Forecasting model initialisation}

We simulate $N_p=10^5$ realisations of the process (referred to as particles). We define an initialisation period $t_\mathrm{init}$ corresponding the largest of $m_g$, $m_r$ and $m_h$. The value of $R_t$ on the last day of the initialisation period (i.e. $t=t_\mathrm{init}$), is drawn from the stationary distribution of the Gaussian process defined in Eq. (5). In other words, $\ln(R_t)$ is drawn from normal distribution with mean $0$ and variance $s_n^2$. The value of $P_t$ at $t=t_\mathrm{init}$ is drawn from a Gamma distribution with mean $\bar{P}$ and coefficient of variation $P_{CV}=0.025$. The value of $\bar{P}$ is estimated directly from the data as the ratio of total hospitalisations to total reported cases over the first $21$ days of the simulation period. 

The number of daily infections $I_t$ on day $t$ during the initialisation period was drawn from an independent Poisson distribution for each particle with mean
\begin{equation}
    i_t = \frac{\tilde{C}_{t+\bar{r}}}{p_c}, \qquad t=1,\ldots,t_\mathrm{init},
\end{equation}
where $\tilde{C}_t$ is the smoothed number of observed cases on day $t$, calculated directly from the case data as a 7-day centred moving average, and $\bar{r}$ is the mean time from infection to reporting. 

Note that the exact method of initialisation for $R_t$, $P_t$ and $I_t$ does not have a substantial effect on model results after an initial burn-in period. The starting time for the simulation was taken to be 250 days prior to the origin date, or the date of first available data, whichever was later.

\subsection{Forecasting model fitting method}
At each time step, the $N_p$ particles were propagated forwards one day by drawing random variables according to Eqs. (5), (7) and (10). Given the observed data on day $t$, the likelihood $L_i$ of each particle $i$ was calculated according to Eqs. (11)--(12). Note that for SARS-CoV-2, data were available for both daily cases ($C_t$) and daily hospitalisations ($A_t$), and so the likelihood of each particle was calculated as the product of the likelihood for $C_t$ and $A_t$ according to Eqs. (11) and (12) respectively. For influenza and RSV, only data on hospitalisations was available, so the likelihood was calculated using Eq. (12) only, with $P_t$ fixed at 1. 

At each time step, particles were sampled with replacement with weights $L_i$ and the process was repeated until the final data point (i.e., the origin date) was reached. To avoid particle degeneracy, we used fixed-lag resampling with a lag of $L=42$ days, meaning that only the most recent 42 days of history were resampled. {\hl This means only marginal posterior distributions of model states are available more than 42 days prior to the most recent data, but does not affect forecasting outputs \cite{watson2024improving}.}
Future trajectories were simulated by propagating the particles forwards from the origin date in the same way, but without any particle resampling. All forecasts were run for 28 days beyond the origin date.

Once we had obtained fitted trajectories for $R_t$, $P_t$, $I_t$, $Z_t$ and $H_t$, we generated samples for the observed variables $C_t$ and $A_t$ according to Eqs. (11)--(12).

\subsection{Parameter values and delay distributions} \label{sec:parameters}

The distributions for the generation interval, time from infection to reporting, and time from infection to hospital admission were assumed to be discretised gamma distributions (parameterised by mean and standard deviation -- see Table 1). 

For each distribution, we truncated the distribution at a specified maximum value and obtained a discretised probability mass function from the continuous Gamma probability density function via the method of \cite{cori2013new}. The minimum generation interval was assumed to be $1$ day; the minimum delay from infection to either reporting or hospital admission was assumed to be $0$ days (although the probability of this occurring was very low). 

The mean and variance of the time from infection to reporting were estimated respectively as the sum of the means and variances of the incubation period and the time from symptom onset to reporting. Similarly for the time from infection to hospital admission. This assumes that the incubation period, the time from symptom onset to reporting and the time from reporting to admission are all independent variables.

Values for the mean and standard deviation of the generation interval and the incubation period were taken from the ACEFA distribution library \cite{acefa2025distribution} for consistency. These were based on estimates in \cite{abbott2022estimation,backer2022shorter,nishiura2011estimation,cauchemez2009household,perofsky2024impacts}. Values for the mean and standard deviation of the time from symptom onset to reporting and the time from reporting to hospital admission were taken from \cite{plank2024near} and assumed to be the same for all three pathogens.

\clearpage
\section{Supplementary results}

\begin{table}[h!]
\centering
\begin{tabular}{lll}
\hline
{\bf Target} & {\bf Coverage (50\% CrI)} & {\bf Coverage (95\% CrI)} \\
\hline
SARS-CoV-2 cases & 60.9\% & 99.5\% \\ 
SARS-CoV-2 hospitalisations & 61.1\% & 97.7\% \\ 
Influenza hospitalisations & 62.7\% & 99.6\% \\ 
RSV hospitalisations & 68.5\% & 98.2\% \\ 
\hline
\end{tabular}
\label{tab:coverage}
\caption{Coverage of the 28-day ahead forecast intervals showing the proportion of data points for each target that fell in the 50\% credible interval and 95\% credible interval.}
\end{table}

\begin{figure}
    \centering
    \includegraphics[width=0.75\linewidth]{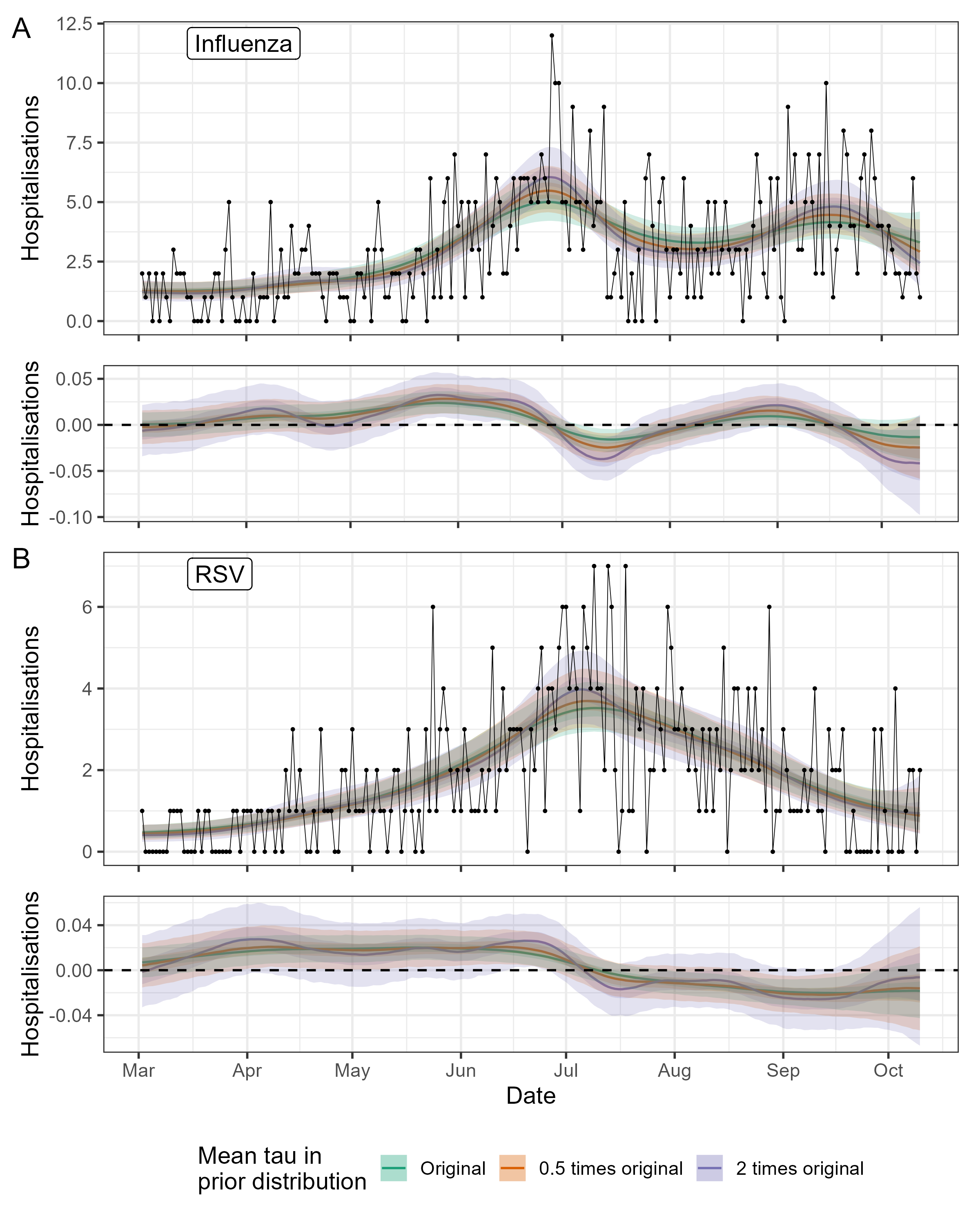}
    \caption{Sensitivity of modelled trends in hospitalisations assuming different prior distributions for the parameter $\tau$ for: (A) influenza; and (B) RSV. In the top panel for each pathogen we plot the daily number of hospitalisations (points) and modelled trend in hospitalisations (line and shaded regions) using a Bayesian P-spline model that does not include a day-of-the-week effect. In the bottom panel for each pathogen we plot the growth rate inferred from the modelled trend in hospitalisations. The dashed line indicates a growth rate of zero (the threshold between epidemic growth or decline). For all panels we include the model posterior distribution's median (line) and 95\% credible intervals (dark shaded and light shaded regions). Three modelled estimates (colours) are shown assuming different prior distributions for the parameter $\tau$ (that controls the smoothness of the second-order random-walk). We include: the original prior distribution used in the main analyses (Green); a prior (normal) distribution with a mean of $0.5\times$the original mean (Orange); and a prior distribution with a mean of $2\times$the original mean (Purple).}
    \label{fig:hosp-trends-overview-sensitivity}
\end{figure}

\begin{figure}
    \centering
    \includegraphics[width=0.75\linewidth]{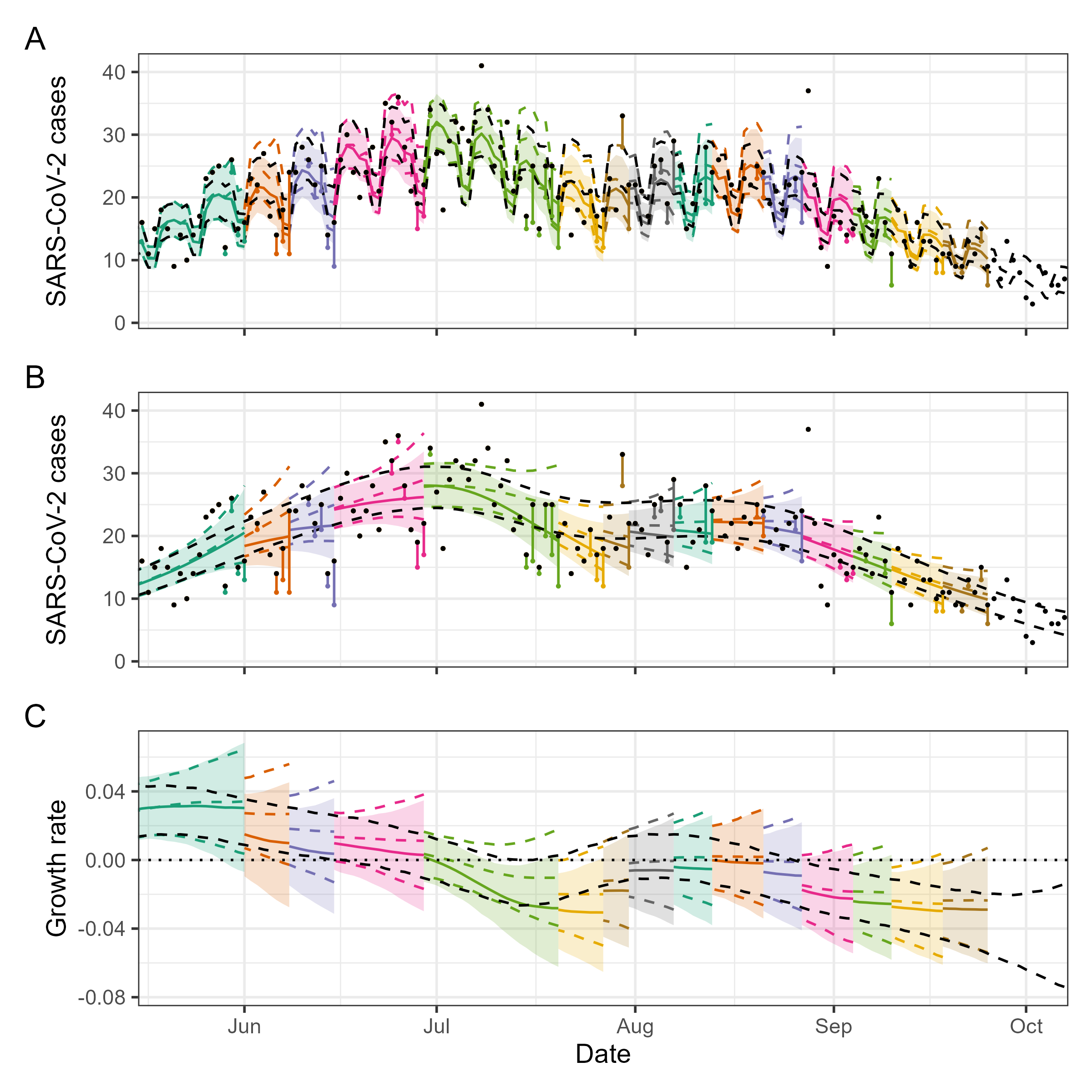}
    \caption{Comparison of modelled trends in SARS-CoV-2 cases estimated using data received in real-time, and estimated using the final dataset (up to the same time points). (A) Modelled trends in SARS-CoV-2 cases including day-of-the-week effect. (B) Modelled trends in SARS-CoV-2 cases with day-of-the-week effect removed (C) Growth rate inferred from the modelled trend in SARS-CoV-2 cases (with day-of-the-week effect removed). For all panels real-time estimates (solid coloured lines and shaded regions) are compared to estimates made over the same time-period but using the final dataset (i.e. removing the effect of data revisions) (solid dashed lines). For all estimates (coloured by round), we include the model posterior distribution's median (solid line, or dashed line) and central 95\% credible intervals (shaded region, or dashed lines). We include the model estimates up to the final day of data (for periods greater than the previous round's final day of data). We also include the 95\% credible intervals of the end-of-season estimates (black dashed lines) using the entire final dataset. In (A) and (B) the daily SARS-CoV-2 cases for the final end-of-season dataset (black points) and the daily SARS-CoV-2 cases in the real-time datasets (points coloured by round) are connected by lines to highlight revisions to data used in real-time analyses (as in Figure 3).
    }
    \label{fig:case-trends-real-time-sensitivity}
\end{figure}

\begin{figure}
    \centering
   \includegraphics[width=0.9\linewidth]{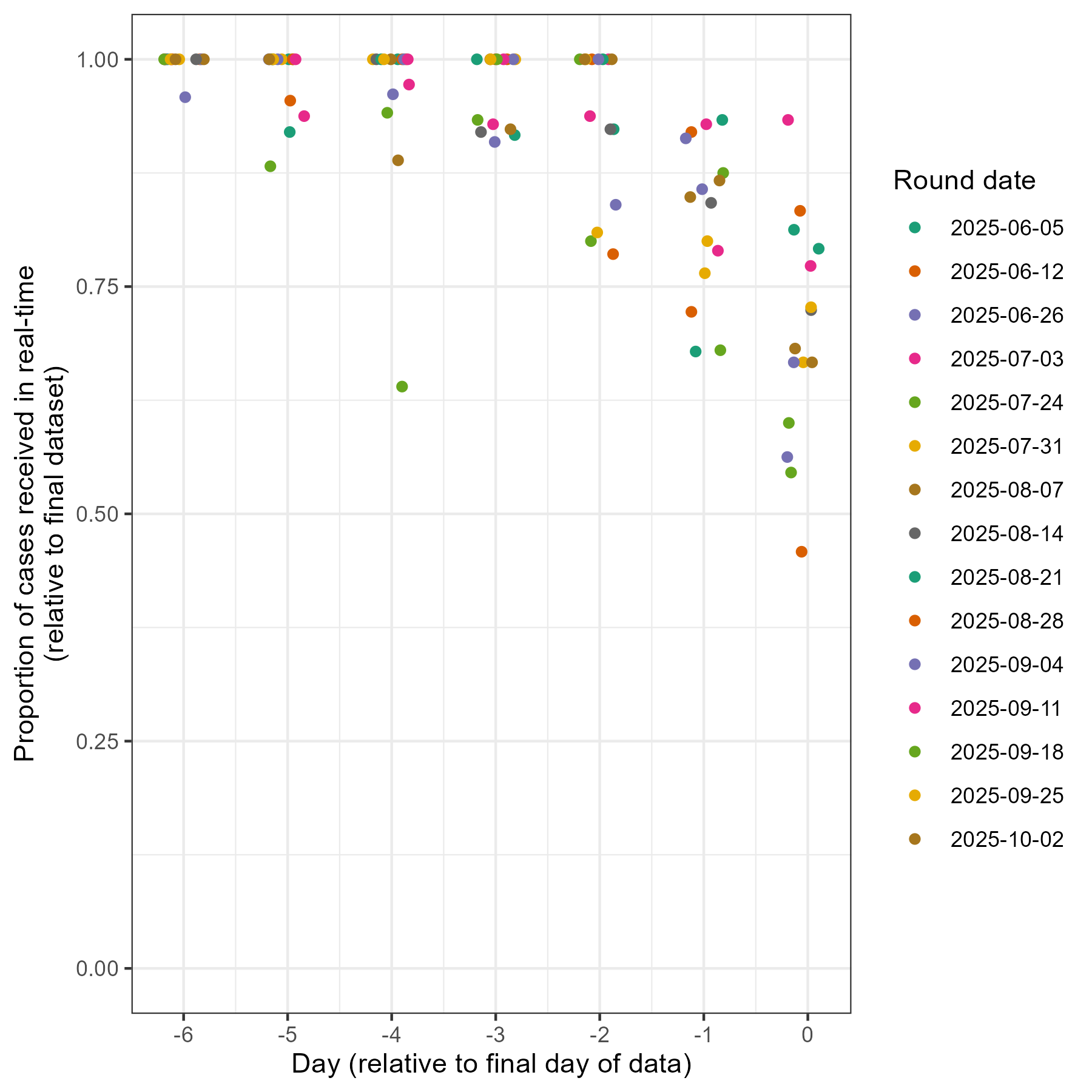}
    \caption{Revisions to SARS-CoV-2 case data received in real-time. The proportion of SARS-CoV-2 cases in the final dataset that were included in datasets received in real-time (coloured points) over the season. The proportions are provided for the final seven days of each dataset received in real-time.}
    \label{fig:data_revisions_cases}
\end{figure}

\begin{figure}
    \centering
   \includegraphics[width=0.8\linewidth]{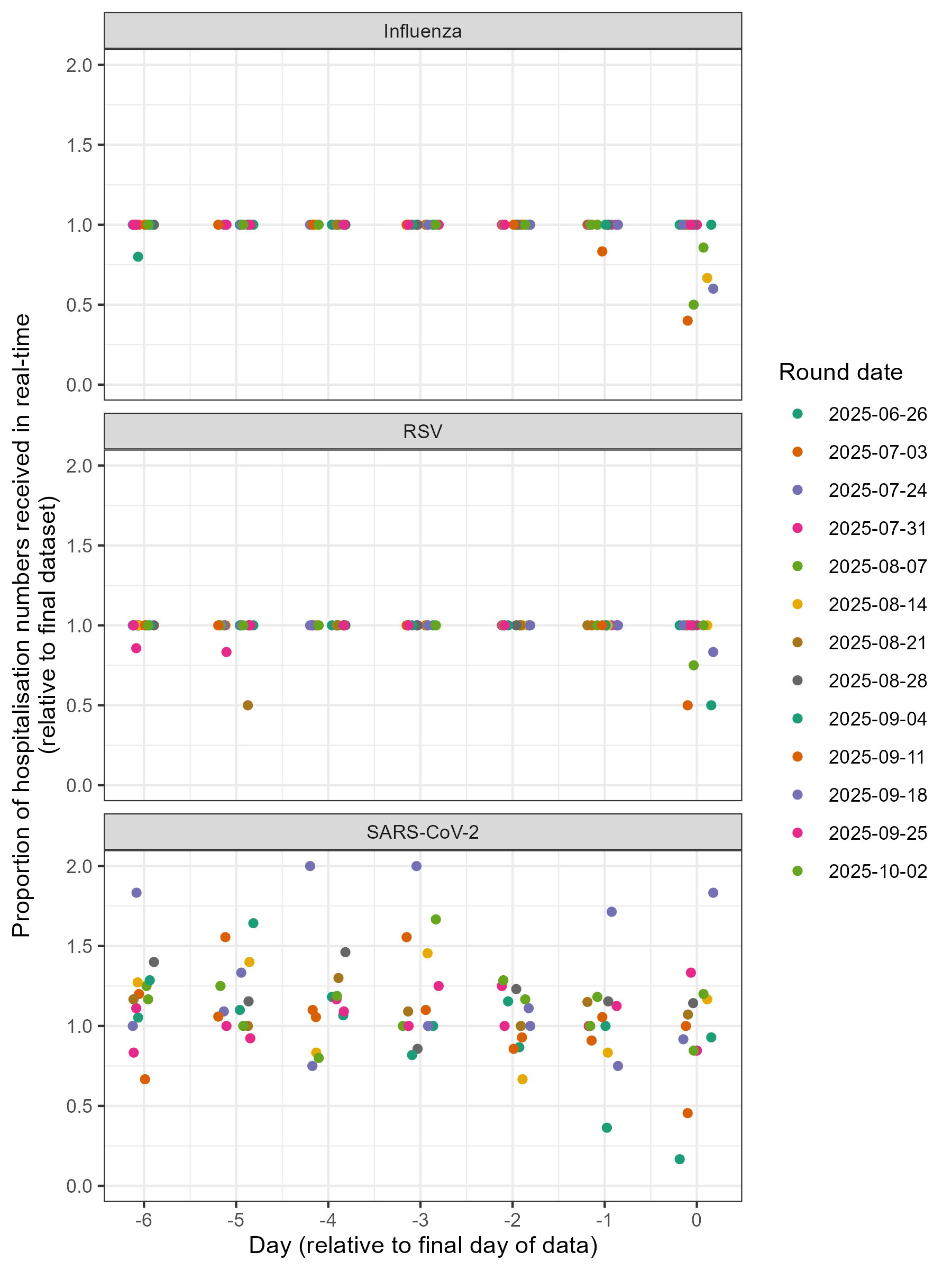}
    \caption{Revisions to hospitalisation data received in real-time. The proportion of hospitalisations in the final dataset that were included in datasets received in real-time (coloured points) over the season for influenza (top), RSV (middle) and SARS-CoV-2 (bottom). The proportions are provided for the final seven days of each dataset received in real-time. For SARS-CoV-2 the proportion can be greater than 1.0; this represents hospitalisations received in real-time being revised downwards in the final dataset.}
    \label{fig:data_revisions_hosps}
\end{figure}

\begin{figure}[h!]
    \centering
    \includegraphics[width=1.1\linewidth]{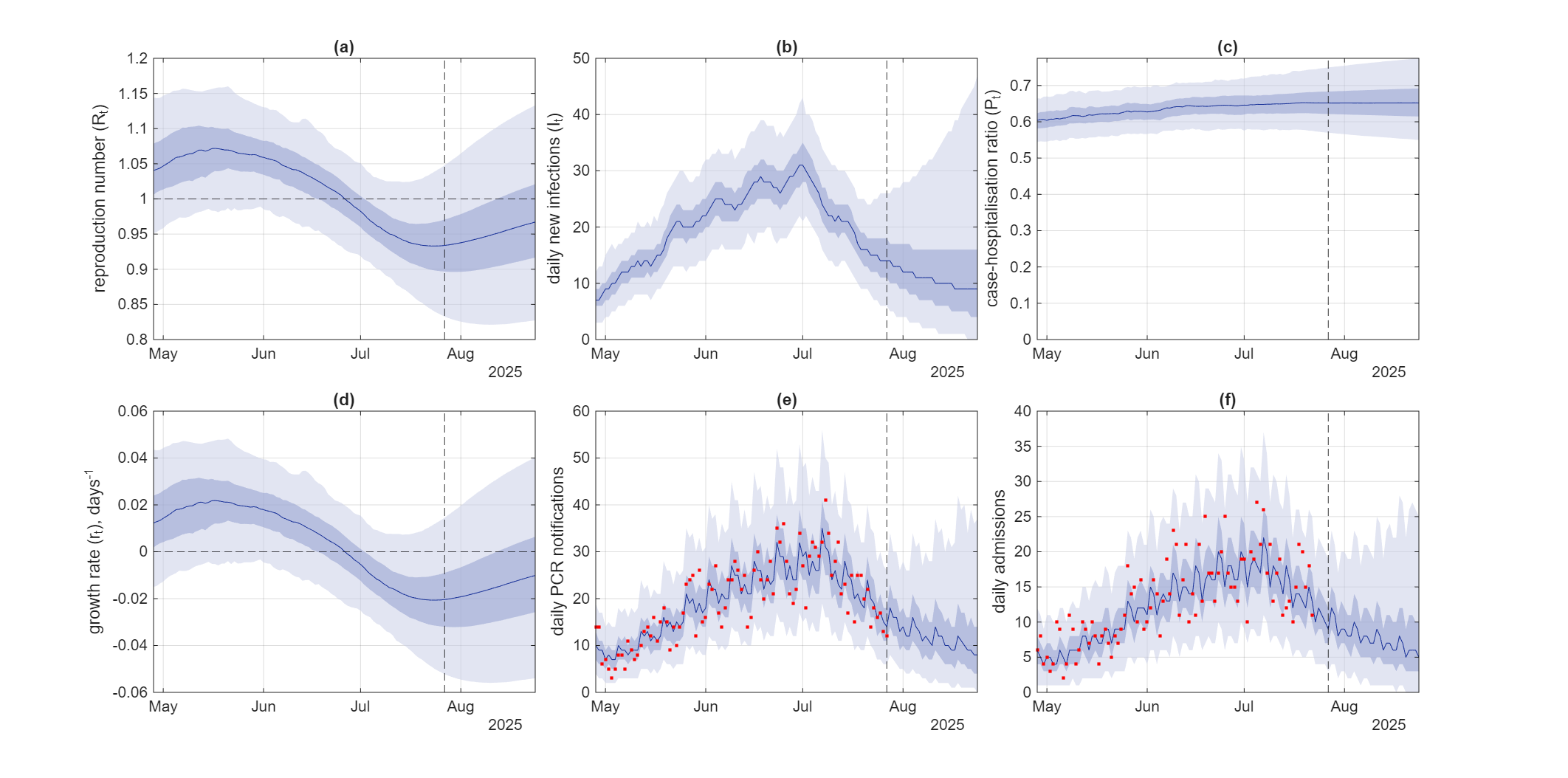}
    \caption{Outputs from the forecasting model for SARS-CoV-2 for an example origin date (27 July 2025, indicated by vertical dashed line) showing: (a)  reproduction number $R_t$; (b) daily incidence of new infections $I_t$; (c) case-hospitalisation ratio $P_t$; (d) epidemic growth rate $r_t$; (e) observed daily number of case notifications $C_t$; (f) observed daily number of new hospital admissions $A_t$. Each plot shows the posterior median (line) and 50\% and 95\% credible intervals (dark shaded and light shaded regions). Red points show observed data.  }
    \label{fig:SARSCOV2_example}
\end{figure}

\begin{figure}[h!]
    \centering
    \includegraphics[width=1.1\linewidth]{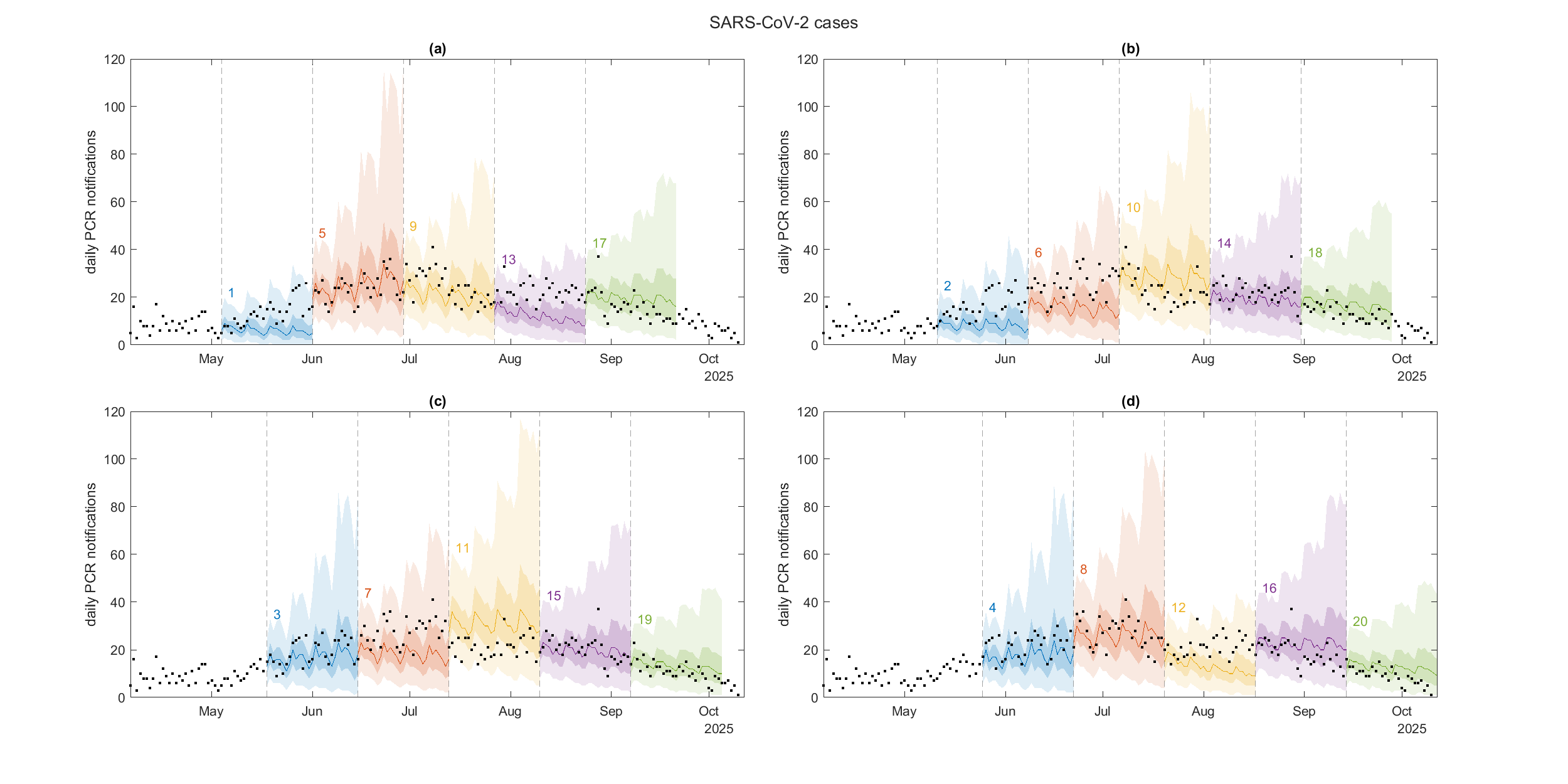}
    \caption{In-season 28-day ahead forecasts for daily SARS-CoV-2 cases at different origin dates (coloured lines and bands) alongside subsequently observed data (black points). Each coloured block in each panel shows a forecast generated using a specific origin date (indicated by the vertical dashed line) labelled according to round number: rounds 1-4 (blue), rounds 5-8 (red), rounds 9-12 (yellow), rounds 13-16 (purple), round 17-20 (green). Each panel shows the same data with forecasts from a different set of origin dates. For each forecast, we show the posterior median (line) and 50\% and 95\% credible intervals (dark shaded and light shaded regions). }
    \label{fig:forecasts_SARSCOV2_cases}
\end{figure}

\begin{figure}[h!]
    \centering
    \includegraphics[width=1.1\linewidth]{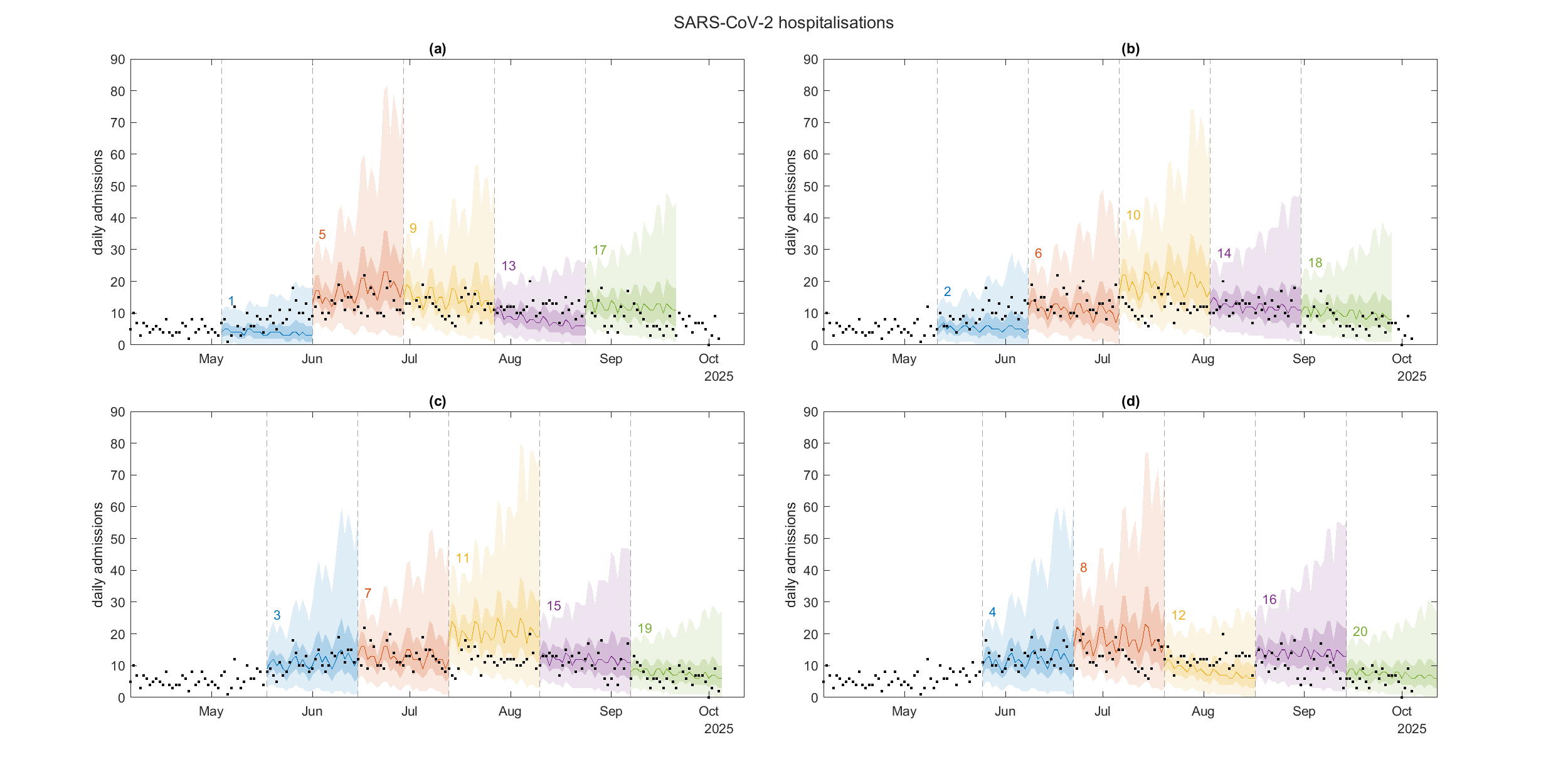}
    \caption{In-season 28-day ahead forecasts for daily SARS-CoV-2 hospitalisations at different origin dates (coloured lines and bands) alongside subsequently observed data (black points). Each coloured block in each panel shows a forecast generated using a specific origin date (indicated by the vertical dashed line) labelled according to round number: rounds 1-4 (blue), rounds 5-8 (red), rounds 9-12 (yellow), rounds 13-16 (purple), round 17-20 (green). Each panel shows the same data with forecasts from a different set of origin dates. For each forecast, we show the posterior median (line) and 50\% and 95\% credible intervals (dark shaded and light shaded regions). }
    \label{fig:forecasts_SARSCOV2_hosp}
\end{figure}

\begin{figure}[h!]
    \centering
    \includegraphics[width=1.1\linewidth]{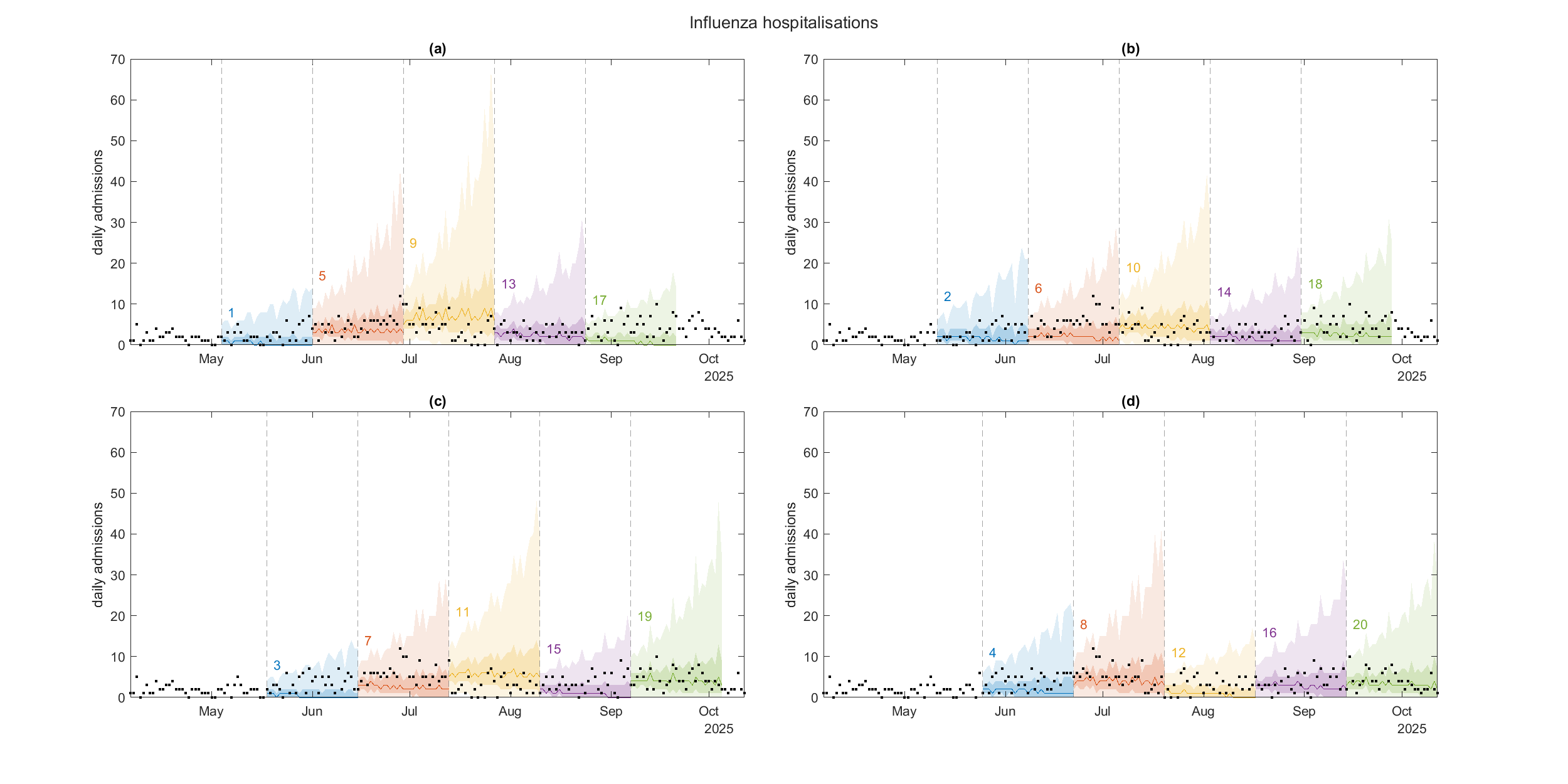}
    \caption{In-season forecasts 28-day ahead for daily influenza hospitalisations at different origin dates (coloured lines and bands) alongside subsequently observed data (black points). Each coloured block in each panel shows a forecast generated using a specific origin date (indicated by the vertical dashed line) labelled according to round number: rounds 1-4 (blue), rounds 5-8 (red), rounds 9-12 (yellow), rounds 13-16 (purple), round 17-20 (green). Each panel shows the same data with forecasts from a different set of origin dates. For each forecast, we show the posterior median (line) and 50\% and 95\% credible intervals (dark shaded and light shaded regions). }
    \label{fig:forecasts_flu_hosp}
\end{figure}

\begin{figure}[h!]
    \centering
    \includegraphics[width=1.1\linewidth]{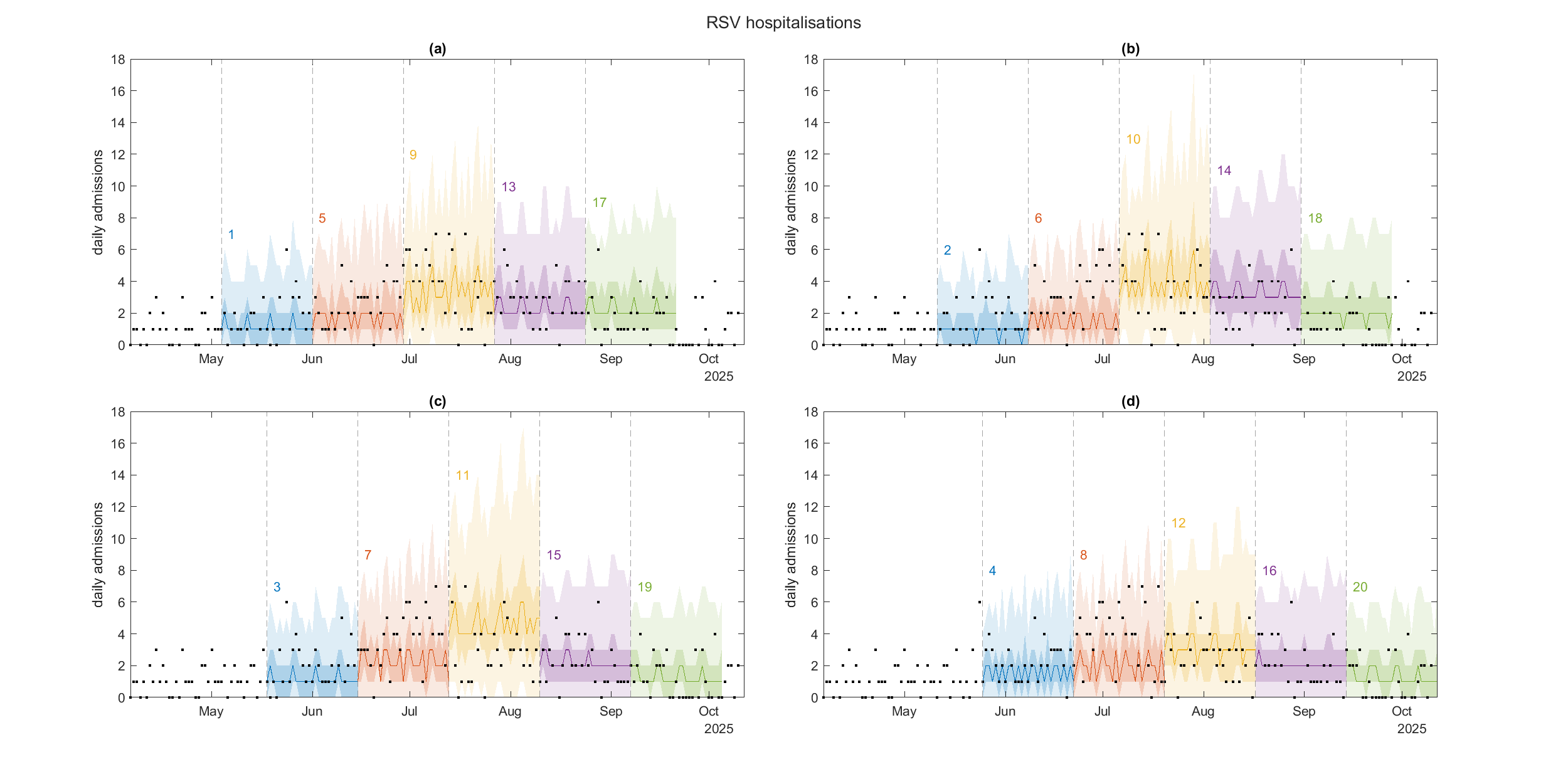}
    \caption{In-season forecasts 28-day ahead for daily RSV hospitalisations at different origin dates (coloured lines and bands) alongside subsequently observed data (black points). Each coloured block in each panel shows a forecast generated using a specific origin date (indicated by the vertical dashed line) labelled according to round number: rounds 1-4 (blue), rounds 5-8 (red), rounds 9-12 (yellow), rounds 13-16 (purple), round 17-20 (green). Each panel shows the same data with forecasts from a different set of origin dates. For each forecast, we show the posterior median (line) and 50\% and 95\% credible intervals (dark shaded and light shaded regions). }
    \label{fig:forecasts_RSV_hosp}
\end{figure}

\clearpage

\bibliography{references}

\clearpage
\section{Example report to public health partners} 
\label{sec:example_report}



\section*{Supplementary material}

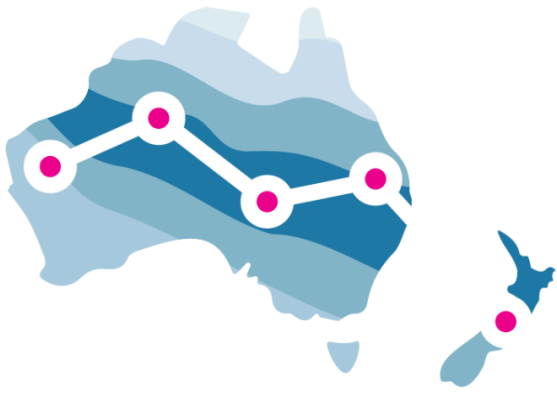

# New Zealand winter respiratory pathogen epidemic analysis

## Report no. 6, 2025

*1 August 2025*

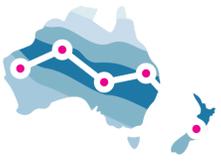

# Contributors

ACEFA involves multiple academic research groups from across Australia and New Zealand, including:

- Infectious Disease Dynamics Unit, The University of Melbourne
- Infectious Disease Ecology and Modelling group, The Kids Institute
- Defence Science and Technology Group
- Monash University
- Infectious Disease Dynamics and Interventions Group, University of New South Wales
- Infectious Disease Epidemiology and Modelling, Australian National University
- Infectious Diseases Modelling and Epidemiology Group, University of Queensland
- Adelaide Data Science Centre, University of Adelaide
- University of Canterbury

---

# Contacts

We welcome any queries, comments, or feedback. Please write to one of our contacts below:

- **Freya Shearer** (Director): [freya.shearer@unimelb.edu.au](mailto:freya.shearer@unimelb.edu.au)
- **Oliver Eales** (Reporting Lead): [oliver.eales@unimelb.edu.au](mailto:oliver.eales@unimelb.edu.au)
- **Michael Plank** (Member of Executive Committee): [michael.plank@canterbury.ac.nz](mailto:michael.plank@canterbury.ac.nz)

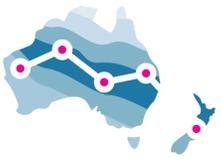

# Overview

This report was produced on 1 August 2025 for distribution to ACEFA public health partners in New Zealand.

The analysis presented uses data provided by PHF Science and Te Whatu Ora Health New Zealand on 31 July 2025.

We present estimates of current trends in SARS-CoV-2, influenza and RSV epidemic activity, including estimates of the growth rate, effective reproduction number, and doubling/halving times through time up to the last day of data. We also provide 28-day ahead forecasts for PCR-confirmed SARS-CoV-2 cases, SARS-CoV-2 hospitalisations, and SARI hospitalisations for influenza and RSV in the Auckland region. We aim to provide additional analyses and assessment as the season progresses. We are also aiming to consolidate analyses across New Zealand and Australia into a single report to provide a more informative picture across the two countries.

---

# Contents

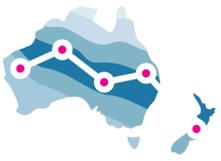

# National picture

<u>SARS-CoV-2</u>
- Following an epidemic peak in late-June/early-July, our analysis suggests that **SARS-CoV-2 cases are highly likely to be decreasing,** but stable epidemic activity is also predicted to be possible (Figure NZ-1, NZ-3, NZ-4). On 27 July:
  - Growth rate was estimated to be -0.03 (-0.07, 0.00) corresponding to a halving time of 23 days (95% CI: halving time of 11 days, doubling time of 600 days).
  - The probability that the growth rate was greater than 0 (cases increasing) was 3%.
  - Effective reproduction number was estimated to be 0.89 (0.79, 0.99).
  - Four-week ahead forecasts of daily case incidence are uncertain, although consistent with the above trends (Figure NZ-3).
- SARS-CoV-2 hospitalisations have increased since late-April, in line with the increase in cases (Figure NZ-2). Our analysis suggests that **SARS-CoV-2 hospitalisations are most likely decreasing or stable**; trends in hospitalisations will likely continue to follow trends in cases. On 22 July:
  - Growth rate was estimated to be -0.01 (-0.04, 0.02) corresponding to a halving time of 63 days (95% CI: halving time of 18 days, doubling time of 42 days).
  - The probability that the growth rate was greater than 0 (cases increasing) was 21%.
  - Four-week ahead forecasts of daily hospital admissions are uncertain, although consistent with the above trends (Figure NZ-4).
- In April 2025 modelled case numbers of SARS-CoV-2 were at the lowest levels observed since January 2023 (Figure NZ-1). SARS-CoV-2 case numbers resurged in New Zealand until late-June/early-July and are now expected to be decreasing following an epidemic peak.
  - Similarly timed resurgences and declines have been observed in Australia (not shown). In all Australian states and the Australian Capital Territory there is evidence that SARS-CoV-2 cases are currently decreasing following recent epidemic peaks.
  - Australian Whole Genome Sequencing samples submitted to GISAID show a recent rapid rise in the proportion belonging to lineage NB.1.8.1 (approximately 7% per day growth advantage estimated in NSW for data up to 14 June, and representing ~80% of reported cases) which rose quickly to dominance in Hong Kong over March and April. Rise of new variants is a potential explanation for recent growth in SARS-CoV-2 cases in several jurisdictions.
  - In 2023 and 2024, broadly similar epidemic timings were observed between New Zealand and some Australian jurisdictions (not shown), likely driven by similar population-level immunity profiles and timing of the introduction of new variants. A resurgence (decline) of SARS-CoV-2 cases in one jurisdiction may be predictive of resurgences (declines) in other jurisdictions.

<u>Influenza</u>
- The expected daily number of influenza hospitalisations has increased since late-March. Currently, our analysis suggests that **influenza hospitalisations are likely decreasing,** but stable epidemic activity is also predicted to be possible. There has been a decrease in the growth rate (from well above to well below zero) since mid-June, suggesting that an epidemic peak occurred in late-June/early-July. On 19 July:
  - Growth rate was estimated to be -0.03 (-0.06, 0.01) corresponding to a halving time of 26 days (95% CI: halving time of 11 days, doubling time of 90 days).
  - The probability that the growth rate was greater than 0 (cases increasing) was 7%.

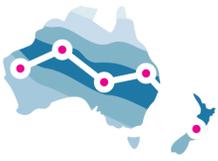

- ○ Recent school holidays may have contributed to a transient decrease in growth. School holidays may have impacted hospitalisation trends due to either changes in transmission or changes in health-care seeking behaviour, or a combination.
- ○ In Australian jurisdictions sharp decreases in influenza cases in school-aged children were observed coinciding (approximately) with the start of each jurisdiction's school holidays. In some jurisdictions a resurgence in school-aged cases have been observed coinciding with the end of school holidays.
- ○ Four-week ahead forecasts of daily hospital admissions are uncertain, but consistent with the above trends (Figure NZ-4)
- ○ The influenza season in Australia has so far been dominated by influenza A H1N1, which currently represents between 63% (QLD) and 88% (the NT) of all influenza cases. Influenza B activity has also increased across all jurisdictions this season and currently represents between 9% (the NT) and 36% (VIC) of all cases. There has been little influenza A H3N2 activity in Australia this season.

RSV

- ● The expected daily number of RSV hospitalisations has increased since early-2025 (first day of data). Currently, our analysis suggests that **RSV hospitalisations are most likely stable,** but decreasing or increasing epidemic activity are also predicted to be possible. On 19 July:
  - ○ Growth rate was estimated to be 0.00 (-0.04, 0.03) corresponding to a halving time of 225 days (95% CI: halving time of 18 days, doubling time of 22 days).
  - ○ The probability that the growth rate was greater than 0 (cases increasing) was 43%.
  - ○ Four-week ahead forecasts of daily RSV hospital admissions are uncertain, but consistent with the above trends (Figure NZ-4)
  - ○ In Australia, trends in modelled estimates of RSV cases (not shown) have been broadly similar to the 2023 and 2024 seasons.


Key methodological notes

- ● We report on epidemic dynamics as inferred from case data or SARI data, noting that these dynamics could reflect changes in the underlying infection dynamics and/or changes in testing and reporting [1].
- ● Analysis of influenza and RSV activity is based on SARI data from hospitals in the Auckland region only, which may not be representative of national trends.
- ● All estimates are reported as median (95% credible interval).
- ● We model dynamics of influenza, RSV and SARS-CoV-2 hospitalisations separately using a single-pathogen Bayesian P-spline model (see Appendix A) that accounts for noise. Note: data on influenza and RSV hospitalisations are PHF Science SARI data from hospitals in the Auckland region; data on SARS-CoV-2 hospitalisations are national COVID-19-related hospital admissions reported to Te Whatu Ora. We do not include the final five days of SARS-CoV-2 data in the hospitalisation analyses to reduce the impact of reporting delays (that would bias our estimates downwards).
- ● We model dynamics of SARS-CoV-2 cases using a single-pathogen Bayesian P-spline model (see Appendix A) that accounts for day-of-the-week effects and other sources of observation noise.
- ● We report estimates of the growth rate and effective reproduction number ($R_t$), noting that $R_t$ estimates are sensitive to the choice of the generation interval distribution [1].

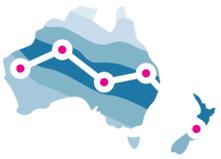
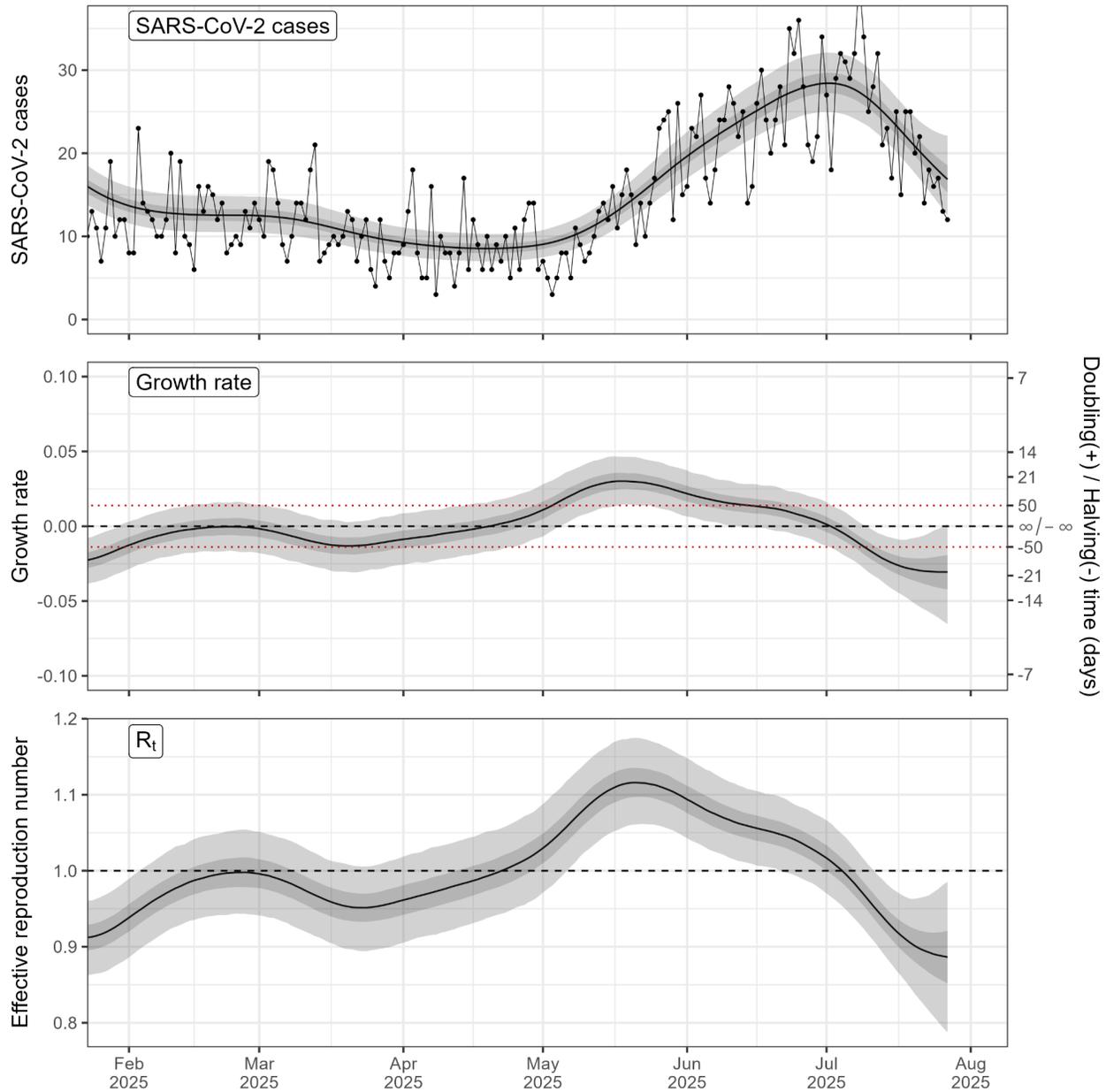

**Figure NZ-1: Modelled dynamics of PCR-confirmed SARS-CoV-2 cases in New Zealand.** (Top panel) Modelled estimates of the expected number of SARS-CoV-2 cases over time. (Middle panel) Modelled estimates of the daily growth rate in the number of SARS-CoV-2 cases. The dashed black line shows where growth rate equals 0, the transition point between epidemic growth and decline. The right-hand y-axis highlights the corresponding values of the doubling (positive)/halving (negative) time, the time taken for the number of cases to reach double/half its current value. The dashed red lines highlight the region in which the doubling/halving time is greater than 50 days and case numbers are approximately stable. (Bottom panel): Modelled estimates of the effective reproduction number of SARS-CoV-2. The dashed line shows where the effective reproduction number equals 1, the transition point between epidemic growth and decline. All estimates are shown over time with median (line), 50% credible intervals (dark shaded region) and 95% credible intervals (light shaded region).

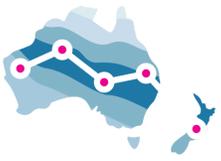

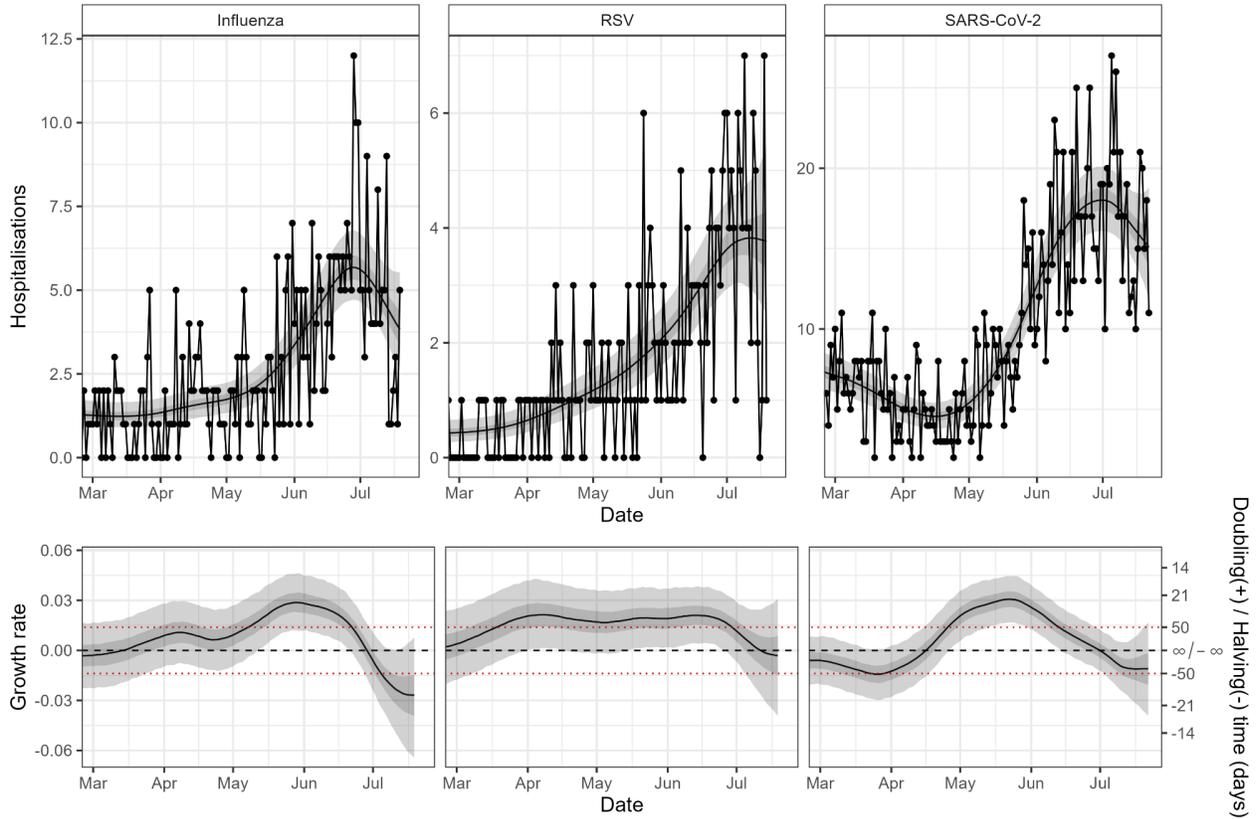

**Figure NZ-2: Modelled dynamics of SARS-CoV-2, RSV, and influenza hospitalisations in New Zealand.** (Top panels) Modelled estimates of the expected number of hospitalisations over time for each pathogen. Note: data on influenza and RSV hospitalisations are PHF Science SARI data from hospitals in the Auckland region; data on SARS-CoV-2 hospitalisations are national Covid-19-related hospital admissions reported to Te Whatu Ora; these graphs should not be used to make comparisons of the number of hospitalisations between pathogens. (Bottom panels) Modelled estimates of the daily growth rate in the number of hospitalisations for each pathogen. The dashed line shows where growth rate equals 0, the transition point between epidemic growth and decline. The right-hand y-axis highlights the corresponding values of the doubling (positive)/halving (negative) time, the time taken for the number of hospitalisations to reach double/half its current value. The dashed red lines highlight the region in which the doubling/halving time is greater than 50 days and hospitalisation numbers are approximately stable. All estimates are shown over time with median (line), 50% credible intervals (dark shaded region) and 95% credible intervals (light shaded region).

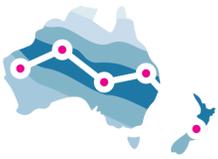

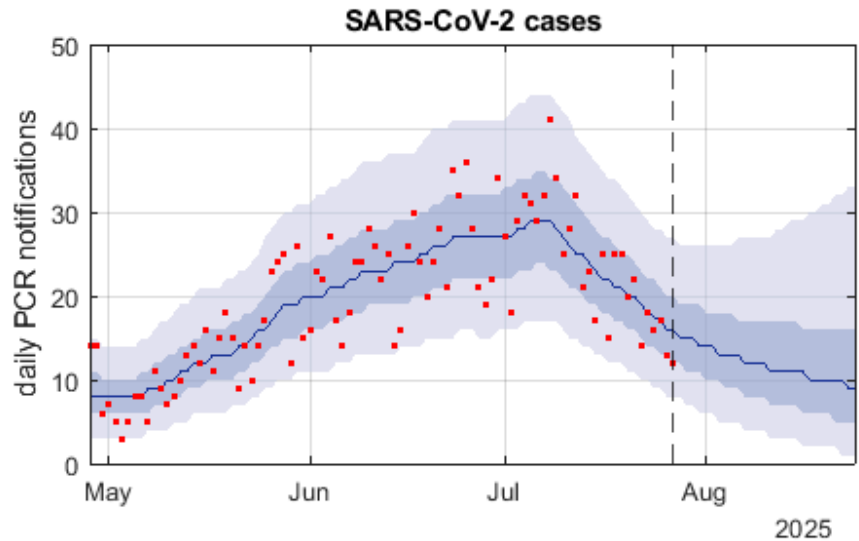

**Figure NZ-3: 28-day ahead forecast for PCR-confirmed SARS-CoV-2 cases in New Zealand**, showing the median (blue curve), 50% credible interval (dark blue band) and 90% credible interval (light blue band). Observed daily case counts are also plotted (red points). Note that the day-of-week effect in the raw data has been quantified and removed from forecast predictions, and so the most recent data point(s) may not fall within the central forecast quantiles. Forecast based on data up to 27 July 2025.

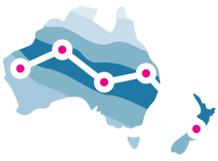

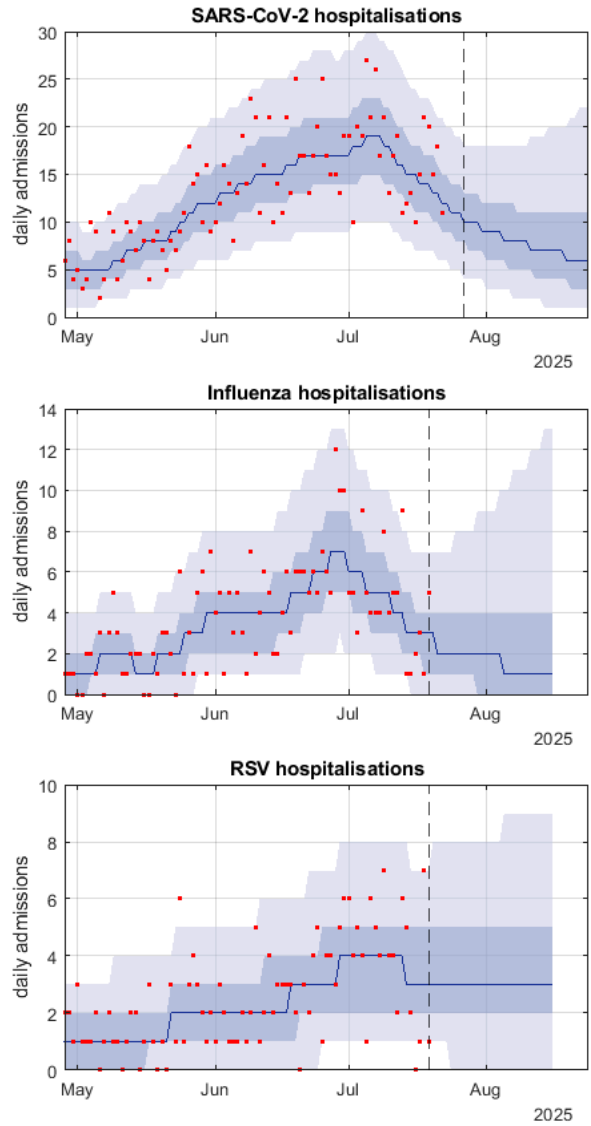

**Figure NZ-4: 28-day ahead forecasts of SARS-CoV-2 hospitalisations, influenza SARI hospitalisations, and RSV SARI hospitalisations in New Zealand.** Note: data on influenza and RSV hospitalisations are PHF Science SARI data from hospitals in the Auckland region; data on SARS-CoV-2 hospitalisations are national Covid-19-related hospital admissions reported to Te Whatu Ora; these graphs should not be used to make comparisons of the number of hospitalisations between pathogens. For all forecasts we show the median (blue curve), 50% credible interval (dark blue band), 90% credible interval (light blue band) and data (red points) for daily hospital admissions. Forecast based on data up to 27 July 2025 for SARS-CoV-2 and 19 July 2025 for influenza and RSV (indicated by the vertical dashed lines).

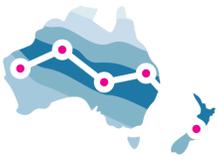

# Appendix A: Bayesian P-spline models

<u>Single-pathogen Bayesian P-spline model</u>

The single-pathogen Bayesian P-spline model has been described in full in [3]. The model has been extended to account for day of the week effects in the case time series.

The model takes as input an epidemic time-series (*e.g.*, the daily number of cases) and estimates the expected value of the epidemic time-series (*i.e.,* a smoothed trend in the daily number of cases accounting for noise). From the smoothed trend we can then estimate the growth rate and effective reproduction number over time (see below).

<u>Multi-pathogen/multi-variant Bayesian P-spline model</u>

The multi-pathogen/multi-variant Bayesian P-spline model is an extension of the two-variant Bayesian P-spline model described in [4]. The model has also been extended to account for day of the week effects in the case time series.

The model takes as input: (1) an epidemic time-series (*e.g.*, the daily number of cases) and (2) data describing the proportion of tests positive for each pathogen/variant/subtype of interest and the number of negative tests. The model then estimates the expected value of the epidemic time-series (*i.e.,* a smoothed trend in the daily number of cases accounting for noise) for each individual pathogen/variant/subtype. From the smoothed trend for each pathogen/variant/subtype we can estimate their growth rate and effective reproduction number over time (see below). We can also estimate the overall trend in the total epidemic time-series (i.e., the sum of all modelled epidemic time-series)

This model is currently only being used to model the individual dynamics of each influenza type/subtype (influenza A H3N2, influenza A H1N1, and influenza B). If data were available on SARS-CoV-2 lineages and variant groupings of concern (i.e., subset of data describing number of test positive for each variant) then we could also estimate individual dynamics for COVID-19 variants. This would also improve estimates of the overall trend in the total number of SARS-CoV-2 cases.

<u>Estimating the growth rates and doubling/halving times</u>

The growth rate, $r(t)$, of an epidemic time-series (e.g., daily counts of cases) is the difference between the log of the modelled case time-series $C(t)$, across two time points
$r(t) = log(C(t)) - log(C(t - 1)).$

The doubling/halving time — the time taken for the modelled value of daily cases to double/half — is then given by the equation:

$doubling\ or\ halving\ time(t) = log(2)/r(t).$

<u>Estimating the effective Reproduction numbers</u>

The effective reproduction number, $R_{eff}(t)$, is the expected number of secondary infections arising from an infected individual. It can be related to the modelled case time-series, $C(t)$, by the equation:

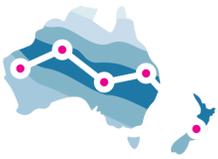

$$R_{eff}(t) = \frac{C(t)}{\int\limits_{0}^{\infty} c(t-\tau)g(\tau)d\tau},$$

where $g(\tau)$ is the intrinsic generation interval, describing the time distribution between primary and secondary infection. This assumes that the modelled epidemic time-series is a reasonable proxy to infection incidence, in reality the epidemic time-series will likely be lagged and smoother compared to actual infection incidence; this could lead to slightly lagged estimates of $R_{eff}(t)$ that may also be biased when the true value of $R_{eff}(t)$ changes rapidly [5].

For estimates of $R_{eff}(t)$ using the modelled estimates from the Bayesian P-spline we have assumed Gamma distributed generation intervals for SARS-CoV-2, RSV, and all influenza subtypes. The parameters of the Gamma distribution used are:

| Pathogen | Shape parameter (gamma distribution) | Rate parameter (gamma distribution) | Mean of distribution | Standard deviation of distribution |
|---|---|---|---|---|
| SARS-CoV-2 (Omicron) [6] | 0.89 | 0.27 | 3.30 | 3.49 |
| RSV [7] | 12.76 | 1.70 | 7.50 | 2.10 |
| Influenza (all subtypes) [8] | 4.00 | 1.54 | 2.60 | 1.30 |

**Table A1. Generation interval distributions used for estimating reproduction numbers**

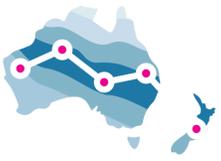

